\documentclass[pdftex,usegraphicx,usenatbib]{mn2e}

\usepackage{graphicx}
\usepackage{natbib}
\usepackage{fixltx2e}
\usepackage{mathptmx}
\usepackage{amsmath}
\usepackage{wasysym}
\usepackage{url}
\usepackage{times}
\usepackage{epsfig}
\usepackage{ulem}

\newcommand{\Mpc}{\rm\; Mpc}
\newcommand{\kpc}{\rm\; kpc}
\newcommand{\pc}{\rm\; pc}
\newcommand{\km}{\rm\; km}
\newcommand{\m}{\rm\; m}
\newcommand{\cm}{\rm\; cm}

\newcommand{\cmpssq}{\hbox{$\cm\s^{-2}\,$}}

\newcommand{\yr}{\rm\; yr}

\newcommand{\Myr}{\rm\; Myr}
\newcommand{\s}{\rm\; s}

\newcommand{\GHz}{\rm\; GHz}

\newcommand{\kHz}{\rm\; kHz}

\newcommand{\K}{\rm\; K}

\newcommand{\g}{\rm\; g}

\newcommand{\gpcmsq}{\hbox{$\g\cm^{-2}\,$}}
\newcommand{\Msun}{\hbox{$\rm\thinspace M_{\odot}$}}

\newcommand{\Msunpsqpc}{\hbox{$\Msun\pc^{-2}\,$}}

\newcommand{\Msunpyr}{\hbox{$\Msun\yr^{-1}\,$}}
\newcommand{\Msunpyrpsqkpc}{\hbox{$\Msunpyr\kpc^{-2}\,$}}

\newcommand{\keV}{\rm\; keV}

\newcommand{\erg}{\rm\; erg}
\newcommand{\Jy}{\rm\; Jy}
\newcommand{\mJy}{\rm\; mJy}

\newcommand{\ergpcmcu}{\hbox{$\erg\cm^{-3}\,$}}

\newcommand{\ergps}{\hbox{$\erg\s^{-1}\,$}}

\newcommand{\Jykmps}{\hbox{$\Jy\km\s^{-1}\,$}}
\newcommand{\Jypbmkmps}{\hbox{$\Jy{\rm /beam}.{\rm km}\s^{-1}\,$}}

\newcommand{\Kkmps}{\hbox{$\K\km\s^{-1}\,$}}

\newcommand{\kmps}{\hbox{$\km\s^{-1}\,$}}

\newcommand{\kmpspMpc}{\hbox{$\kmps\Mpc^{-1}\,$}}

\newcommand{\Lsun}{\hbox{$\rm\thinspace L_{\odot}$}}

\newcommand{\Zsun}{\hbox{$\thinspace \mathrm{Z}_{\odot}$}}

\newcommand{\muG}{\hbox{$\rm\thinspace {\mu}\mathrm{G}$}}
\newcommand{\asec}{\rm\; arcsec}

\newcommand{\psqcm}{\hbox{$\cm^{-2}\,$}}
\newcommand{\pcmsq}{\hbox{$\cm^{-2}\,$}}

\newcommand{\COtoH}{\hbox{$\psqcm(\K\kmps)^{-1}$}}

\begin{document}


\title[ALMA observations of PKS\,0745-191]{ALMA observations of cold
  molecular gas filaments trailing rising radio bubbles in
  PKS\,0745-191}\author[H.R. Russell et al.]
{\parbox[]{7.in}{H.~R. Russell$^{1}$\thanks{E-mail:
      hrr27@ast.cam.ac.uk}, B.~R. McNamara$^{2,3}$, A.~C. Fabian$^{1}$, P.~E.~J. Nulsen$^{4,5}$, A.~C. Edge$^{6}$, F. Combes$^{7}$, N.~W. Murray$^8$, I.~J. Parrish$^8$, P. Salom\'e$^7$, J.~S. Sanders$^9$, S.~A. Baum$^{10}$, M. Donahue$^{11}$, R.~A. Main$^8$, R.~W. O'Connell$^{12}$, C.~P. O'Dea$^{10}$, J.~B.~R. Oonk$^{13,14}$, G. Tremblay$^{15}$, A.~N. Vantyghem$^2$, G.~M. Voit$^{11}$ \\
    \footnotesize
    $^1$ Institute of Astronomy, Madingley Road, Cambridge CB3 0HA\\
    $^2$ Department of Physics and Astronomy, University of Waterloo, Waterloo, ON N2L 3G1, Canada\\
    $^3$ Perimeter Institute for Theoretical Physics, Waterloo, Canada\\
    $^4$ Harvard-Smithsonian Center for Astrophysics, 60 Garden Street, Cambridge, MA 02138, USA\\
    $^5$ ICRAR, University of Western Australia, 35 Stirling Hwy, Crawley, WA 6009, Australia\\ 
    $^6$ Department of Physics, Durham University, Durham DH1 3LE\\
    $^7$ Observatoire de Paris, LERMA, 61 Av. de l'Observatoire, 75014 Paris, France\\
    $^8$ Canadian Institute for Theoretical Astrophysics, University of Toronto, 60 St. George Street, Toronto, M5S 3H8 ON, Canada\\
    $^9$ Max-Planck-Institut f\"ur extraterrestrische Physik, Giessenbachstrasse 1, D-85748 Garching, Germany\\
    $^{10}$ University of Manitoba, Department of Physics and Astronomy, Winnipeg, MB R3T 2N2, Canada\\
    $^{11}$ Department of Physics and Astronomy, Michigan State University, 567 Wilson Road, East Lansing, MI 48824, USA\\
    $^{12}$ Department of Astronomy, University of Virginia, PO Box 400235, Charlottesville, VA 22904, USA\\
    $^{13}$ Netherlands Institute for Radio Astronomy (ASTRON), Postbus 2, NL-7990 AA Dwingeloo, the Netherlands\\
    $^{14}$ Leiden Observatory, Leiden University, PO Box 9513, NL-2300 RA Leiden, the Netherlands\\
    $^{15}$ Department of Physics and Yale Center for Astronomy \& Astrophysics, Yale University, 217 Prospect Street, New Haven, CT 06511, USA\\
  } }

\maketitle

\begin{abstract}
We present ALMA observations of the CO(1-0) and CO(3-2) line emission tracing filaments of cold molecular gas in the central galaxy of the cluster PKS\,0745-191.  The total molecular gas mass of $4.6\pm0.3\times10^{9}\Msun$, assuming a Galactic $X_{\mathrm{CO}}$ factor, is divided roughly equally between three filaments each extending radially $3-5\kpc$ from the galaxy centre.  The emission peak is located in the SE filament $\sim1\asec$ ($2\kpc$) from the nucleus.  The velocities of the molecular clouds in the filaments are low, lying within $\pm100\kmps$ of the galaxy's systemic velocity.  Their FWHMs are less than $150\kmps$, which is significantly below the stellar velocity dispersion.  Although the molecular mass of each filament is comparable to a rich spiral galaxy, such low velocities show that the filaments are transient and the clouds would disperse on $<10^7\yr$ timescales unless supported, likely by the indirect effect of magnetic fields.  The velocity structure is inconsistent with a merger origin or gravitational free-fall of cooling gas in this massive central galaxy.  If the molecular clouds originated in gas cooling even a few kpc from their current locations their velocities would exceed those observed.  Instead, the projection of the N and SE filaments underneath X-ray cavities suggests they formed in the updraft behind bubbles buoyantly rising through the cluster atmosphere.  Direct uplift of the dense gas by the radio bubbles appears to require an implausibly high coupling efficiency.  The filaments are coincident with low temperature X-ray gas, bright optical line emission and dust lanes indicating that the molecular gas could have formed from lifted warmer gas that cooled in situ.
\end{abstract}




\section{Introduction}
\label{sec:intro}


The cores of rich galaxy clusters host the brightest and most massive
galaxies known.  These brightest cluster galaxies (BCGs) are giant
ellipticals with extended, diffuse stellar envelopes and predominantly
old, `red and dead' stellar populations.  However, not all BCGs are
passively evolving at late times.  Those residing beneath cooling hot
atmospheres commonly feature luminous, filamentary emission line
nebulae, cold molecular gas reservoirs with masses above $10^9\Msun$
to a few $\times10^{10}\Msun$ and star formation rates at several to
tens of solar masses per year (\citealt{Heckman81}; \citealt{Hu85};
\citealt{Johnstone87}; \citealt{McNamara92}; \citealt{Jaffe97};
\citealt{Edge01}; \citealt{Salome03}).  Cool core galaxy clusters have
hot X-ray atmospheres with short central radiative cooling times,
which can drop below a Gyr.  An unimpeded flow of gas cooling from the
cluster atmosphere would be expected to supply at least an order of
magnitude more cold gas and star formation than is observed in BCGs in
cool core clusters (\citealt{Fabian94}; \citealt{PetersonFabian06}).
Instead, high resolution X-ray images from \textit{Chandra} show that
powerful radio jets, launched by the central radio AGN, are inflating
large, buoyant radio bubbles and generating shocks, sound waves and
turbulence which heat the cluster atmosphere and suppress gas cooling
(eg. \citealt{McNamaraNulsen07,McNamara12}; \citealt{Fabian12}).  The
energy input by the AGN is sufficient to replace the X-ray radiative
losses for large samples of groups and clusters and appears to be
closely coupled to the cooling rate in a feedback loop
(\citealt{Birzan04}; \citealt{DunnFabian06}; \citealt{Rafferty06}).
This mechanism is likely operating to suppress gas cooling and star
formation in all massive elliptical galaxies at late times
(\citealt{Bower06}; \citealt{Croton06}; \citealt{Best07}).

This radio-mode (mechanical) feedback prevents the bulk of the volume-filling hot
atmospheres in galaxies and clusters from cooling to low temperatures, but whether it can also regulate
the supply of dense, cold molecular gas is not known.  Cold gas and
recent star formation in BCGs are likely fuelled by gas cooling from
the cluster atmosphere, albeit at rates well below the expectations of
cooling flows.  Correlations between the X-ray cooling rate and star
formation rates (\citealt{Egami06}; \citealt{ODea08}) and the
detection of star formation and luminous emission line nebulae
predominantly in systems with central cooling times below a sharp
threshold at $\sim5\times10^8\yr$ support this picture
(\citealt{Rafferty08}; \citealt{Cavagnolo08}).  Accretion of this cold
gas is likely a key element of feedback linking the gas cooling rate
to the fuelling of the SMBH and the energy output of the AGN
(eg. \citealt{PizzolatoSoker05}; \citealt{Gaspari15}).  Radio-jet
driven outflows of ionized and molecular gas have been detected in
nearby radio galaxies suggesting that jets can couple to dense gas
clouds (\citealt{Morganti05}; \citealt{Nesvadba06};
\citealt{Alatalo11}; \citealt{Dasyra11}; \citealt{Morganti15}).  In
NGC\,1275 at the centre of the Perseus cluster, the velocity structure
of extended H$\alpha$ filaments, which are coincident with single dish
detections of molecular gas, is consistent with uplift under the
buoyantly rising radio bubbles (\citealt{HatchPer06};
\citealt{Salome06,Salome11}).  Inflowing molecular gas filaments are
also observed closer to the galaxy centre (\citealt{Lim08}).  ALMA
Early Science observations of Abell 1835 showed a much more
substantial uplift with $10^{10}\Msun$ of molecular gas in a high
velocity flow underneath the buoyant radio bubbles
(\citealt{McNamara14}).

Here we present ALMA Cycle 1 observations of the molecular gas in the
PKS\,0745-191 BCG, which with $P_{\mathrm{cav}}\sim5\times10^{45}\ergps$ has
undergone an even more powerful radio jet outburst than Abell 1835
(\citealt{Rafferty06}; \citealt{Sanders14}).  Single dish observations
of PKS\,0745-191 detected the BCG at CO(1-0) and CO(2-1) and found a total
molecular gas mass of $3.8\pm0.9\times10^{9}\Msun$
(\citealt{Salome03}).  Optical, UV and IR observations show
significant star formation at a rate of $\sim20\Msunpyr$ and a luminous
emission line nebula with clumpy filaments extending $>10\kpc$ in the
BCG (\citealt{Fabian85}; \citealt{Johnstone87}; \citealt{Donahue00}; \citealt{HicksMushotzky05};
\citealt{Tremblay15}).  The surrounding rich cluster has a short
central cooling time $<5\times10^{8}\yr$ and X-ray spectra are
consistent with several hundred solar masses per year cooling down
below X-ray temperatures.  The coolest X-ray gas is offset by
$\sim2.5\asec$ ($\sim5\kpc$) to the W of the hard X-ray and radio
point source emission from the low luminosity AGN, which may be due to
sloshing of the hot gas in the cluster potential
(\citealt{Sanders14}).  The ALMA observations now resolve the spatial
and velocity structure of the cold molecular gas structures at the
centre of the BCG revealing extended filaments trailing the two
buoyant radio bubbles. 


For a standard $\Lambda$CDM cosmology with $H_0=70\kmpspMpc$, at the
central galaxy's redshift ($z=0.1028$; \citealt{Hunstead78}) 1 arcsec
corresponds to $1.9\kpc$.  The BCG redshift was determined from
optical emission lines, including H$\alpha$, H$\beta$ and
[O~\textsc{iii}], which likely originate from the ionized surfaces of
the molecular gas clouds making up the extended cool gas nebula (eg. \citealt{Jaffe05}; \citealt{HatchPer06}; \citealt{Oonk10}; \citealt{Salome11}).  Bulk
motion of the emission line nebula could therefore produce a
systematic offset in the gas velocities with respect to the
gravitational potential of the BCG.  In the absence of stellar
absorption line measurements directly tracing the BCG potential, we
note that our conclusions on the velocity structure of the molecular
gas depend on the redshift from the emission line gas.

\section{Data reduction}
\label{sec:reduction}

\begin{figure}
\centering
\includegraphics[width=0.95\columnwidth]{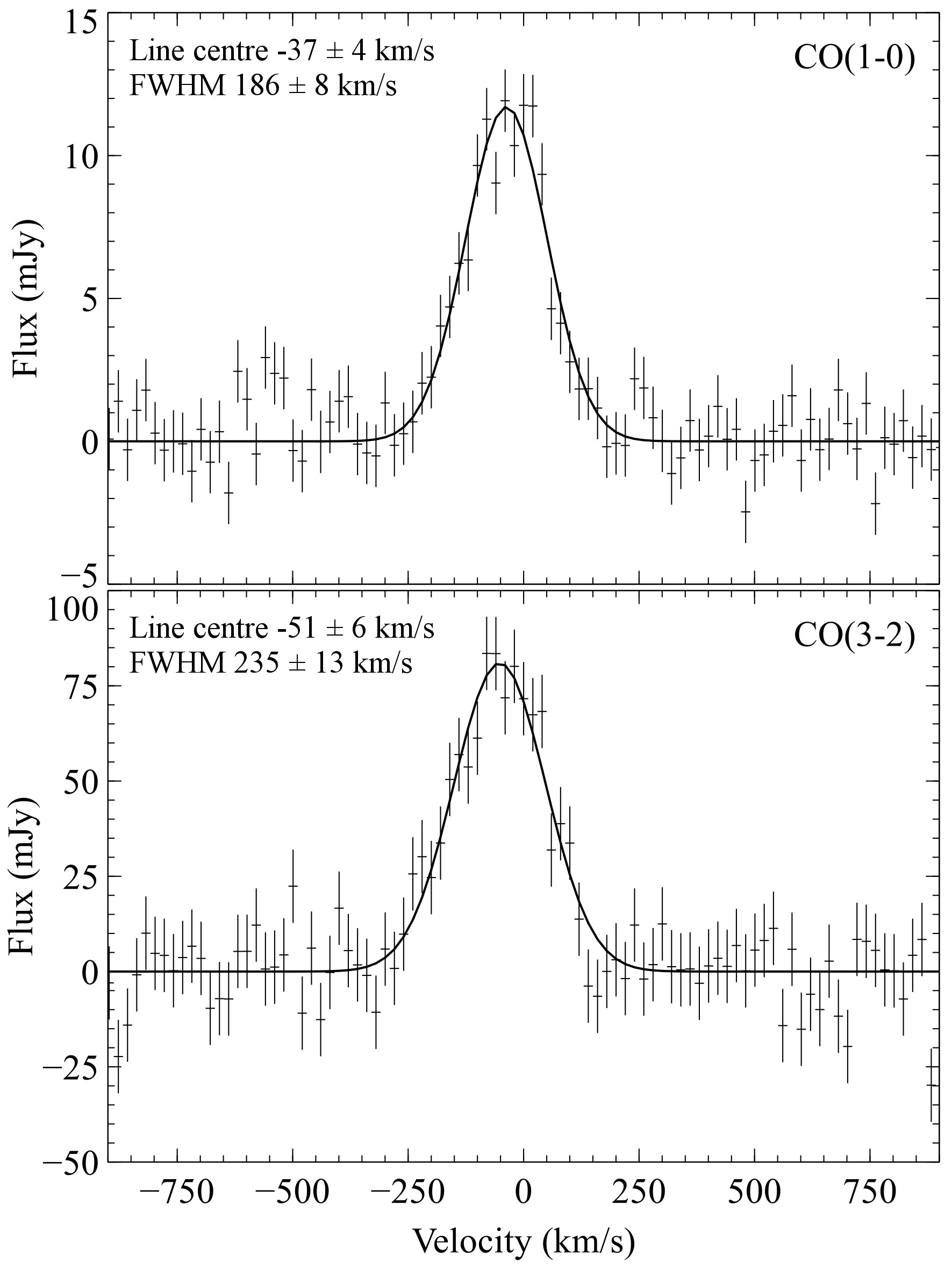}
\caption{PKS\,0745-191 CO(1-0) (top) and CO(3-2) (bottom) total spectra
  for $6\arcsec\times6\arcsec$ and $4\arcsec\times4\arcsec$ regions,
  respectively.  The best-fit model parameters are given in Table \ref{tab:fits}.}
\label{fig:totalspec}
\end{figure}

\begin{table*}
\begin{minipage}{\textwidth}
\caption{Fit parameters for the CO(1-0) and CO(3-2) spectra from different regions.}
\begin{center}
\begin{tabular}{l c c c c c c c c c}
\hline
CO & $\nu_{\rm rest}$ & $\nu_{\rm obs}$ & Region & $\chi^2$/dof & Integrated intensity & Peak & FWHM & Velocity shift & Mass\footnote{The molecular gas mass was calculated from the CO(1-0) integrated intensity as described in section \ref{sec:mass}.} \\
 line & (GHz) & (GHz) & & & (Jy{\thinspace}km/s) & (mJy) & (km/s) & (km/s) & ($10^{9}M_{\odot}$) \\
\hline
J=1-0 & 115.27 & 104.53 & Total & 1079/1061 & $2.2\pm0.1$ & $10.9\pm0.4$ & $186\pm8$ & $-37\pm4$ & $4.6\pm0.3$ \\
 & & & SE & 1072/1058 & $0.49\pm0.04$ & $2.4\pm0.1$ & $190\pm10$ & $-29\pm5$ & \\
 & & & & & $0.06\pm0.02$ & $1.0\pm0.2$ & $50\pm10$ & $-212\pm5$ & \\
 & & & N & 1077/1058 & $0.43\pm0.04$ & $2.5\pm0.1$ & $160\pm10$ & $-87\pm6$ & \\
 & & & & & $0.04\pm0.02$ & $1.0\pm0.3$ & $40\pm20$ & $28\pm6$ & \\
 & & & SW & 1070/1061 & $0.45\pm0.04$ & $2.7\pm0.1$ & $160\pm10$ & $-27\pm4$ & \\
J=3-2 & 345.80 & 313.56 & Total & 350/357 & $18.6\pm1.4$ & $74\pm4$ & $235\pm13$ & $-51\pm6$ & \\
 & & & Nuc. & 332/354 & $0.7\pm0.2$ & $1.8\pm0.4$ & $390\pm70$ & $-130\pm30$ & \\
 & & & & & $0.26\pm0.09$ & $2.9\pm0.6$ & $80\pm20$ & $-68\pm8$ & \\
 & & & SW1 & 374/357 & $0.97\pm0.07$ & $8.0\pm0.4$ & $114\pm6$ & $-6\pm3$ & \\
 & & & SW2 & 340/354 & $0.58\pm0.08$ & $7.3\pm0.5$ & $75\pm9$ & $-39\pm4$ & \\
 & & & & & $0.17\pm0.09$ & $1.9\pm0.5$ & $90\pm40$ & $60\pm20$ & \\
 & & & SE1 & 371/357 & $1.62\pm0.09$ & $7.1\pm0.3$ & $214\pm9$ & $-24\pm4$ & \\
 & & & SE2 & 351/354 & $1.8\pm0.4$ & $5\pm1$ & $310\pm30$ & $-10\pm10$ & \\
 & & & & & $0.8\pm0.2$ & $6\pm1$ & $130\pm20$ & $-49\pm5$ & \\ 
 & & & N1 & 401/357 & $0.53\pm0.08$ & $3.7\pm0.4$ & $130\pm20$ & $-142\pm6$ & \\
 & & & N2 & 372/354 & $0.37\pm0.07$ & $3.1\pm0.4$ & $110\pm20$ & $-161\pm7$ & \\
 & & & & & $0.18\pm0.05$ & $2.9\pm0.5$ & $60\pm10$ & $3\pm5$ & \\
 & & & N3 & 382/354 & $0.34\pm0.07$ & $4.3\pm0.6$ & $80\pm10$ & $-151\pm6$ & \\
 & & & & & $0.2\pm0.1$ & $1.8\pm0.4$ & $130\pm50$ & $-30\pm20$ & \\
 & & & N4 & 371/357 & $0.37\pm0.06$ & $4.6\pm0.4$ & $77\pm9$ & $-145\pm4$ & \\
 & & & N5 & 346/357 & $0.28\pm0.06$ & $3.6\pm0.5$ & $70\pm10$ & $-69\pm5$ & \\
\hline
\end{tabular}
\end{center}
\label{tab:fits}
\end{minipage}
\end{table*}


The BCG at the centre of PKS\,0745-191 was observed by ALMA in Cycle 1
with single pointings to cover the CO(1-0) line at $104.53\GHz$ in band 3 and
the CO(3-2) line at $313.56\GHz$ in band 7 (ID = 2012.1.00837.S; PI
McNamara).  The observations were centred on the nucleus of PKS\,0745-191 (RA 07:47:31.32, Dec -19:17:39.97, J2000)
and the HPBW of the primary beam was $60\asec$ in band 3 and $20\asec$ in band 7.  The data
were taken in two 8 minute observations in band 3 on 2014 April 26 and
27 and in a 25 minute observation in band 7 on 2014 August 19.  The
observations utilised $32-36$ antennas with baselines of $20-560\m$ at
CO(1-0) and a more extended configuration with baselines of
$34-1100\m$ at CO(3-2).  The frequency division correlator mode was
used with a $1.875\GHz$ bandwidth and frequency resolution of $488.3\kHz$
($1.40-0.47\kmps$) but channels were binned together to improve the
signal-to-noise ratio.  An additional baseband was positioned to image
the continuum in each observation.  

The observations were calibrated in \textsc{casa} version 4.2.2
(\citealt{McMullin07}) using the ALMA pipeline reduction scripts and
additional self-calibration significantly improved the image quality.
The continuum-subtracted images were reconstructed using
\textsc{clean} and various Briggs weightings were explored to
determine the optimum for imaging in each band.  A robustness
parameter of 0.5 was used for the CO(1-0) cube and a value of 2 was
used to maximise the signal-to-noise in the CO(3-2) cube.  This
provided a synthesized beam of $1.6\arcsec\times1.2\arcsec$ with a
position angle (P.A.) of $79.7^{\circ}$ at CO(1-0) and
$0.27\arcsec\times0.19\arcsec$ with a P.A. of $78.3^{\circ}$ at
CO(3-2).  The rms noise in the line-free channels was $0.6\mJy$ at
CO(1-0) and $1\mJy$ at CO(3-2) for $10\kmps$ channels.  Images of the
continuum emission were also produced by averaging channels free of
any line emission.  An unresolved central continuum source is detected
in both bands with flux $9.30\pm0.04\mJy$ at $103.8\GHz$ and
$4.6\pm0.2\mJy$ at $314.8\GHz$.  The position of the mm-continuum
source coincides with the unresolved radio nucleus detected at
$8.64\GHz$ with the ATCA in 6A configuration (\citealt{Hogan15}).  The
mm-continuum flux is also consistent with the upper limit on
synchrotron emission from a flat spectrum radio core\footnote{For the
  convention $f_{\nu} \propto \nu^{-\alpha}$} with $\alpha=0.2$
(\citealt{Hogan15}) and this is likely to be the location of the
low-luminosity AGN.

\section{Results}

\subsection{Gas distribution in the central galaxy}

The CO(1-0) and CO(3-2) rotational transition lines were both detected
and imaged at the centre of the BCG in PKS\,0745-191.  The
continuum-subtracted total spectral line profiles were extracted from
a $6\asec\times6\asec$ region for CO(1-0) and a $4\asec\times4\asec$
region for CO(3-2) and are shown in Fig. \ref{fig:totalspec}.  Larger
regions produce consistent total fluxes but significantly noisier
spectra.  Each spectrum was fitted with a single Gaussian component
using the package \textsc{mpfit} (\citealt{Markwardt09}) and the best
fit results, corrected for primary beam response and instrumental
broadening, are shown in Table \ref{tab:fits}.  The CO(1-0) and
CO(3-2) total spectra are consistent with single velocity components
blueshifted to $-30$ to $-50\kmps$ and have a comparable FWHM of
$200-230\kmps$.  The total CO(1-0) integrated intensity of
$2.5\pm0.2\Jykmps$ is roughly consistent with the IRAM $30\m$ single
dish signal of $1.8\pm0.4\Jykmps$ (\citealt{Salome03}) given the
significant uncertainties in the continuum baseline subtraction for
the earlier detection.  This shows that little extended
emission has been filtered out by the interferometer at CO(1-0).

\begin{figure}
\centering
\includegraphics[width=0.98\columnwidth]{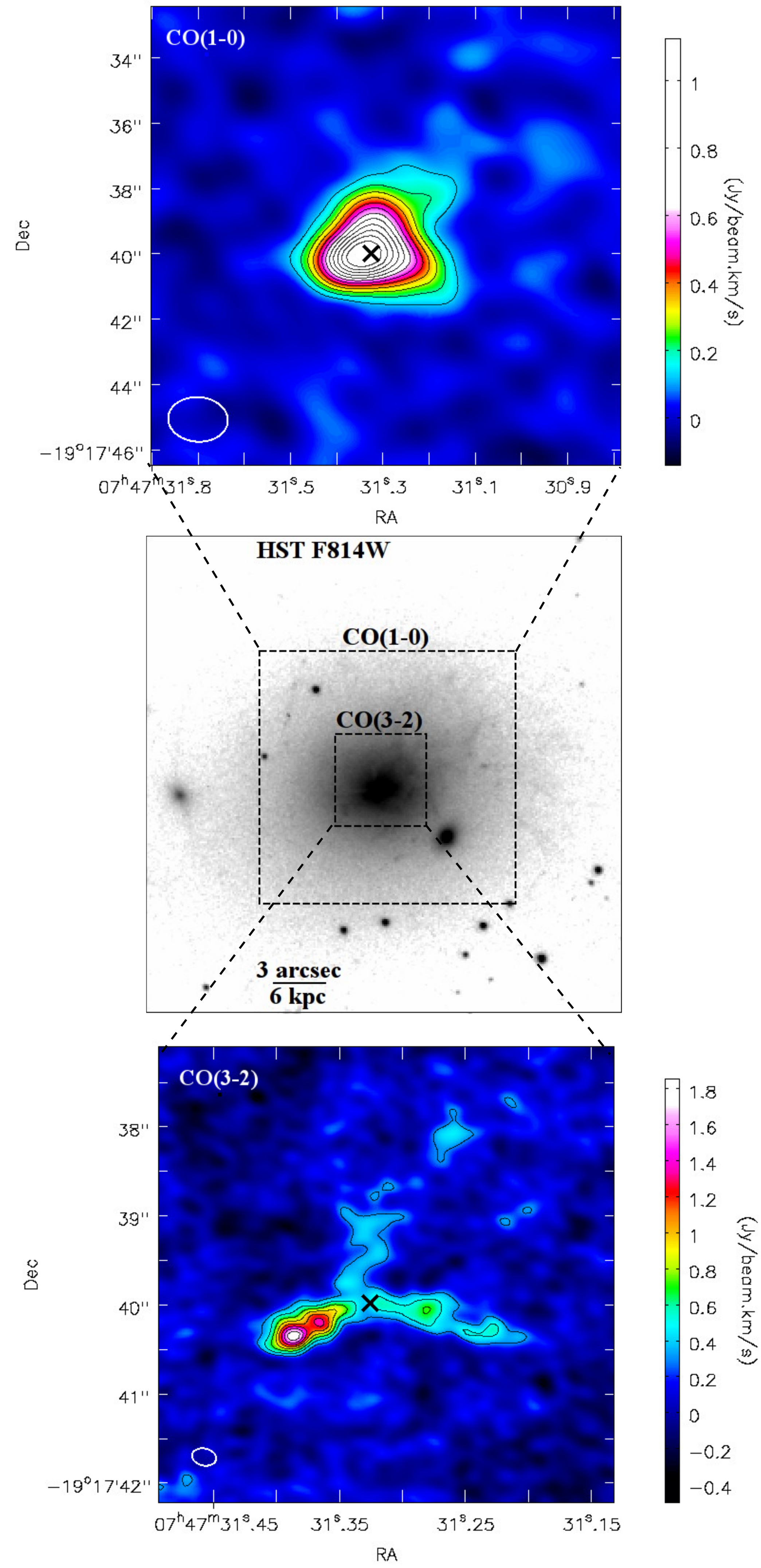}
\caption{Top: CO(1-0) integrated intensity map
  for velocities $-240$ to $160\kmps$.  Centre: HST F814W image of the BCG at the centre of PKS\,0745-191
  with regions corresponding to the size of the ALMA images overlaid.  The image includes the H$\alpha{+}$[N\textsc{ii}] line emission and the old stellar population with a  half light radius of $\sim20\kpc$ (2MASS extended source K-band; \citealt{Skrutskie06}). Bottom: CO(3-2) integrated
  intensity map for velocities $-240$ to $180\kmps$.  Contours are at
  $-3\sigma$, $3\sigma$, $5\sigma$, $7\sigma\dots$, where
  $\sigma=0.04\Jypbmkmps$ for CO(1-0) and $\sigma=0.114\Jypbmkmps$ for
  CO(3-2).  The ALMA beam size is shown lower left and the continuum
  point source location is shown by the black cross.}
\label{fig:COtotalintensity}
\end{figure}

Integrated intensity maps of the CO(1-0) and CO(3-2) emission are
shown in Fig. \ref{fig:COtotalintensity}.  The CO(1-0) line emission
extends over $\sim6\asec$ ($11\kpc$) with roughly a third of the
emission lying within an unresolved central peak that is marginally
offset to the SE of the nuclear continuum emission.  The CO(1-0)
emission extends to the N, SE and SW of the nucleus with a similar
morphology to the brightest regions of H$\alpha$, Pa$\alpha$ and
ro-vibrational H$_2$ (\citealt{Donahue00}; \citealt{Wilman09}).  The
CO(3-2) line emission resolves this extended structure into three
filaments each $1.5-2.5\asec$ ($3-5\kpc$) in length.  The SE, SW and N
filaments contain roughly $35\%$, $30\%$ and $25\%$ of the total line
flux in CO(3-2), respectively.  The N filament extends furthest and
appears to bend along its length and may be fragmenting at large
radius.  The peak in the CO(3-2) emission lies $1.1\asec$ to the SE of
the nuclear continuum emission.  Emission from the SW filament appears
to extend across the nucleus and this structure may also include the
base of the N filament.  However, several structures are likely
superimposed at this junction between the three filaments.  The
molecular gas is clearly not centred on the nucleus and it is
difficult to constrain the physical location of the structure along the line
of sight.



\subsection{Line ratio}
\label{sec:lineratios}

The CO(3-2)/CO(1-0) line ratio was calculated for each of the three
filaments and a region centred on the nucleus.  The CO(3-2) dataset
was convolved with a 2D Gaussian to match the CO(1-0) synthesized beam
and the integrated intensities (in $\Kkmps$) were determined from
spectral fitting to each dataset using matching CO(1-0) beam-sized
regions.  The line ratios for the BCG centre, N and SE filaments were
consistent within the error with CO(3-2)/CO(1-0)$=0.81\pm0.05$,
$0.8\pm0.1$ and $0.9\pm0.1$, respectively.  The line ratio measured
for the SW filament was significantly lower at
CO(3-2)/CO(1-0)$=0.53\pm0.07$.  The FWHM measured in each of the four
regions is consistent within the errors for both the CO(1-0) and the
matched resolution CO(3-2) observation.  Assuming that the CO emission
is optically thick and the gas is thermalized, the line ratios
indicate an excitation temperature of $20-30\K$.  The gas is therefore
highly excited and dense.  The lower line ratio for the SW filament is
likely due to a lower average molecular density in this region and
more diffuse gas.

\subsection{Velocity structure}

We mapped the velocity structure of the molecular gas by extracting
spectra for synthesized beam-sized regions centred on each spatial
pixel in the cube.  These spectra were fitted with one or two Gaussian
components using \textsc{mpfit} and we required at least $3\sigma$
significance for the detection of a line determined using Monte Carlo
simulations.  As discussed in section \ref{sec:intro}, the BCG's
systemic velocity has been determined from optical emission line
spectra and may therefore be affected by the bulk motion of the
emission line gas with respect to the BCG potential.  However, the
radial distribution of the filaments in PKS\,0745-191 appears very
different from the one-sided bulk offsets of the emission line gas
from the BCG centre in systems undergoing mergers or gas sloshing
(\citealt{Hamer12}).  We therefore do not expect a large velocity
offset $>100\kmps$ between the gas structures and the stellar
potential.  

\begin{figure*}
\begin{minipage}{\textwidth}
\centering
\includegraphics[width=0.45\columnwidth]{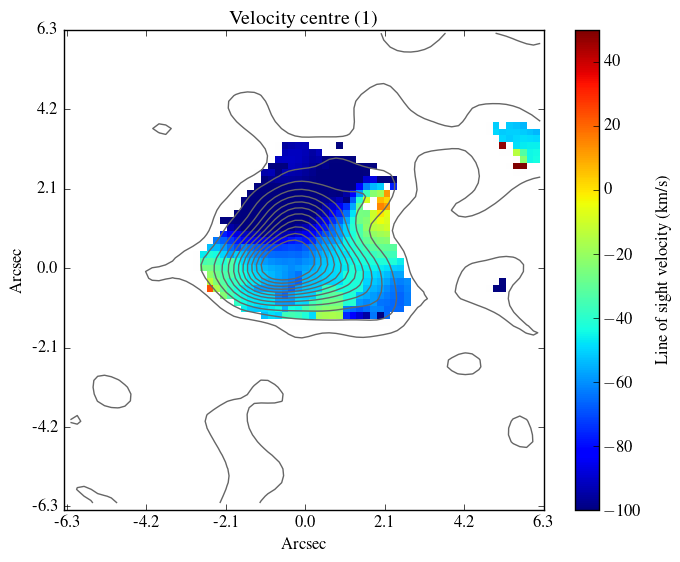}
\includegraphics[width=0.45\columnwidth]{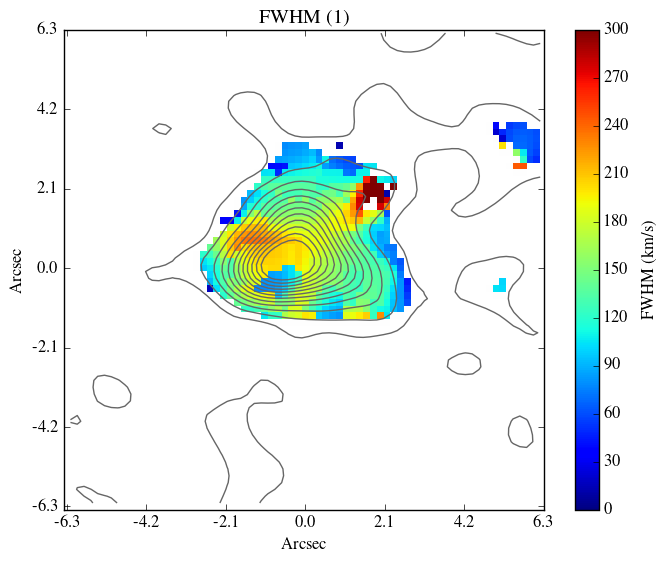}
\includegraphics[width=0.45\columnwidth]{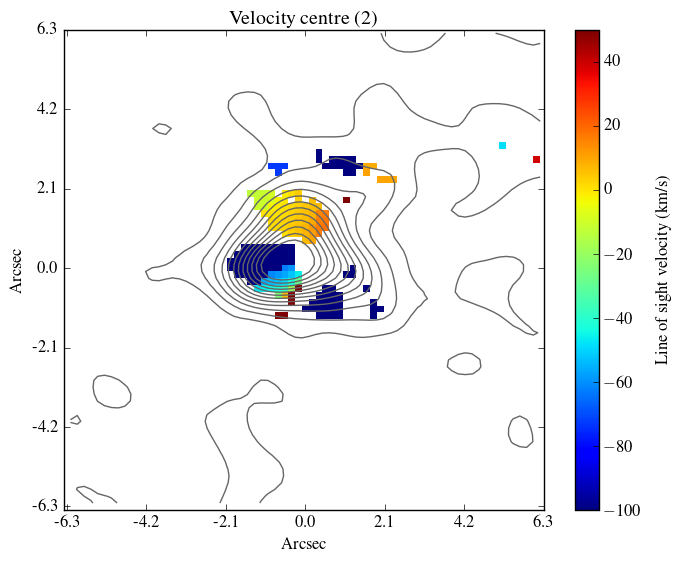}
\includegraphics[width=0.45\columnwidth]{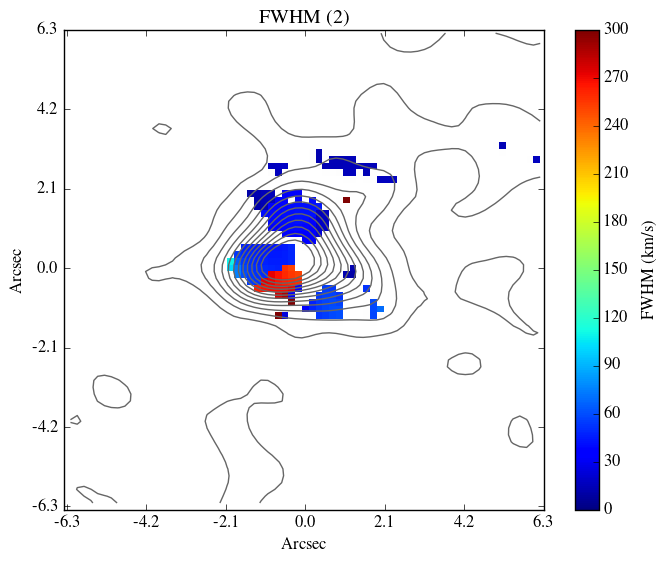}
\caption{Maps of the CO(1-0) velocity centre (left) and FWHM (right)
  for each velocity component (1 and 2, where components with similar
  velocity centres have been grouped together).  The image is centred
  on the nuclear continuum emission (section \ref{sec:reduction}) and contours from the integrated
  CO(1-0) intensity map at $1\sigma, 3\sigma, 5\sigma\dots$ are shown
  overplotted.  Spectra for the regions to the N and SE of the nucleus
  that require two velocity components are shown in
  Fig. \ref{fig:CO102compspec}.}
\label{fig:CO10vmaps}
\end{minipage}
\end{figure*}

\begin{figure*}
\begin{minipage}{\textwidth}
\centering
\includegraphics[width=0.45\columnwidth]{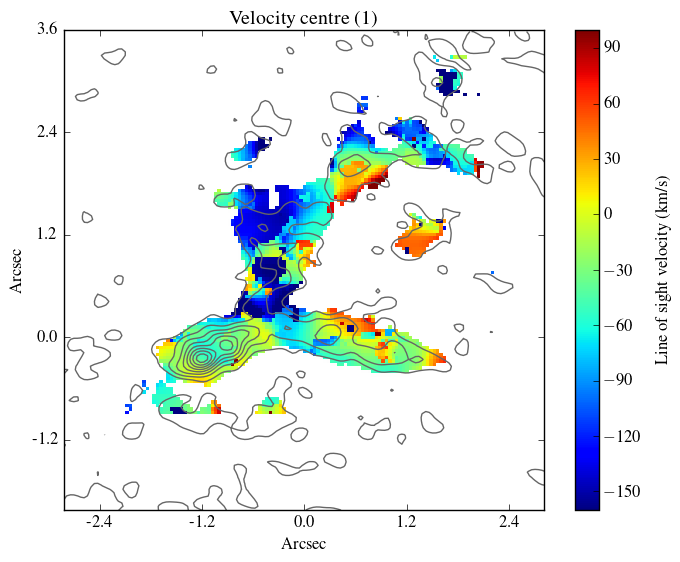}
\includegraphics[width=0.45\columnwidth]{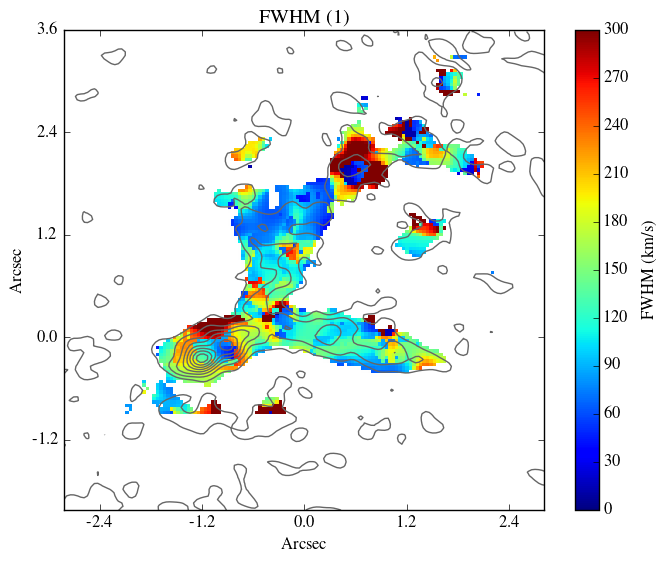}
\includegraphics[width=0.45\columnwidth]{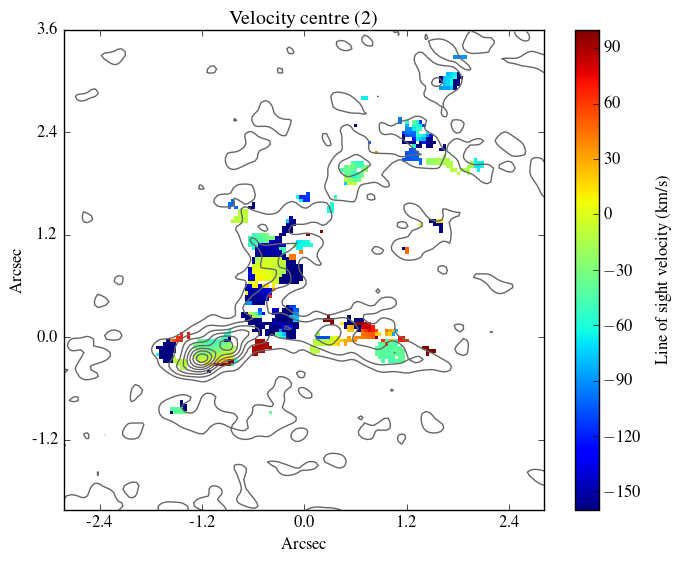}
\includegraphics[width=0.45\columnwidth]{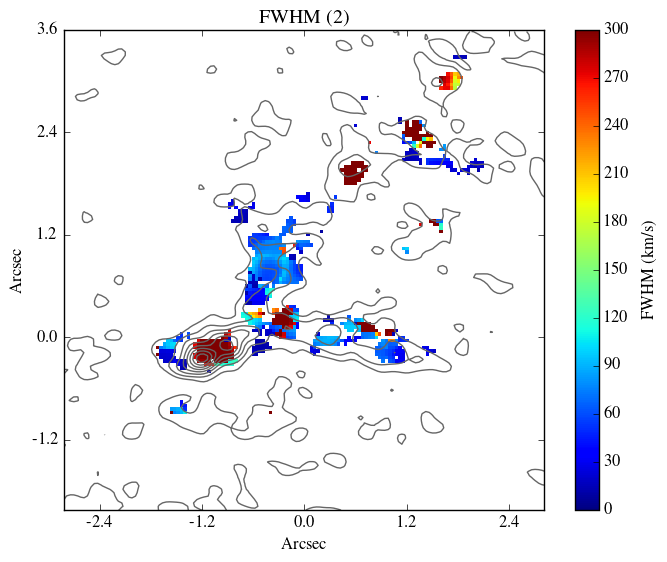}
\caption{Maps of the CO(3-2) velocity centre (left) and FWHM (right) for each velocity component (1 and 2, where components with similar
  velocity centres have been grouped together).  The origin is located at the position of the nuclear continuum emission (section \ref{sec:reduction}) and contours from the integrated CO(3-2) intensity map at $1\sigma, 3\sigma, 5\sigma\dots$ are shown overplotted.  Spectra for regions in each filament requiring two velocity components are shown in Fig. \ref{fig:CO32compspec}.}
\label{fig:CO32vmaps}
\end{minipage}
\end{figure*}

\begin{figure}
\centering
\includegraphics[width=0.98\columnwidth]{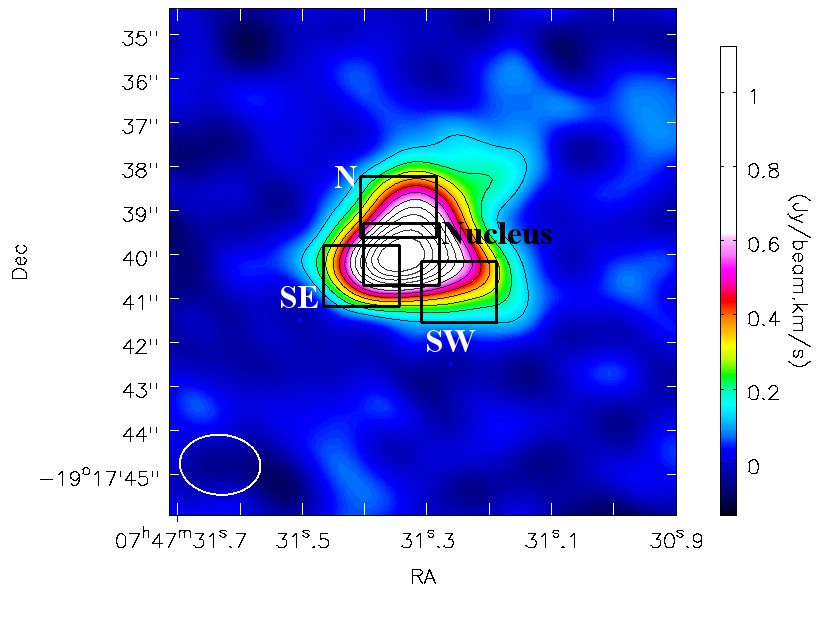}
\includegraphics[width=0.98\columnwidth]{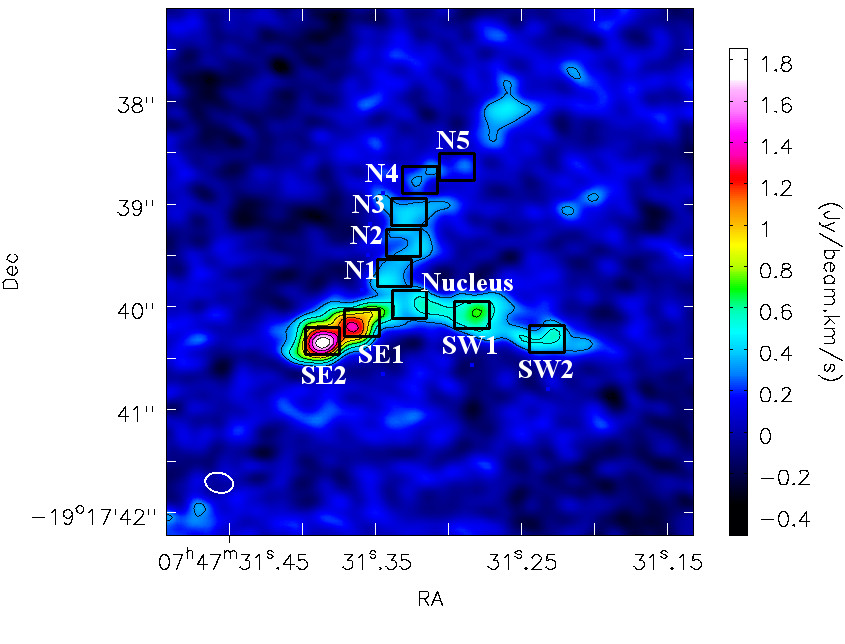}
\caption{PKS\,0745-191 CO(1-0) (top) and CO(3-2) (bottom) integrated intensity maps (as shown in Fig. \ref{fig:COtotalintensity}) with regions overlaid corresponding to the spectra shown in Figs. \ref{fig:CO102compspec} and \ref{fig:CO32compspec}.}
\label{fig:COtotalintensitywregs}
\end{figure}

\begin{figure}
\centering
\includegraphics[width=0.95\columnwidth]{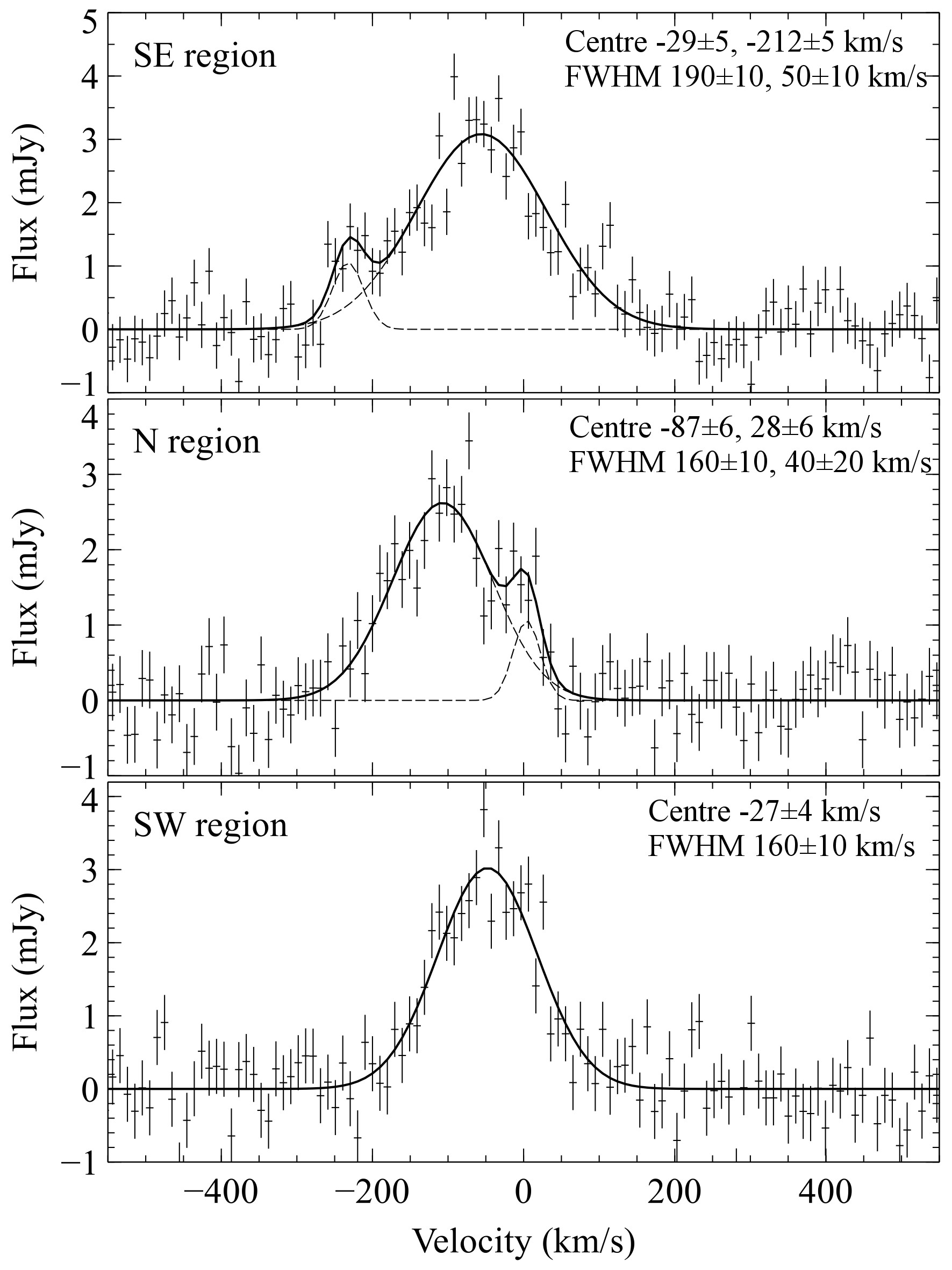}
\caption{CO(1-0) spectra from regions requiring two velocity components and a comparison region to the SW.  The best-fit model is shown by the solid line and individual Gaussian components are shown by the dashed lines.  The best-fit parameters are given in Table \ref{tab:fits}.}
\label{fig:CO102compspec}
\end{figure}

\begin{figure*}
\begin{minipage}{\textwidth}
\centering
\includegraphics[width=0.7\columnwidth]{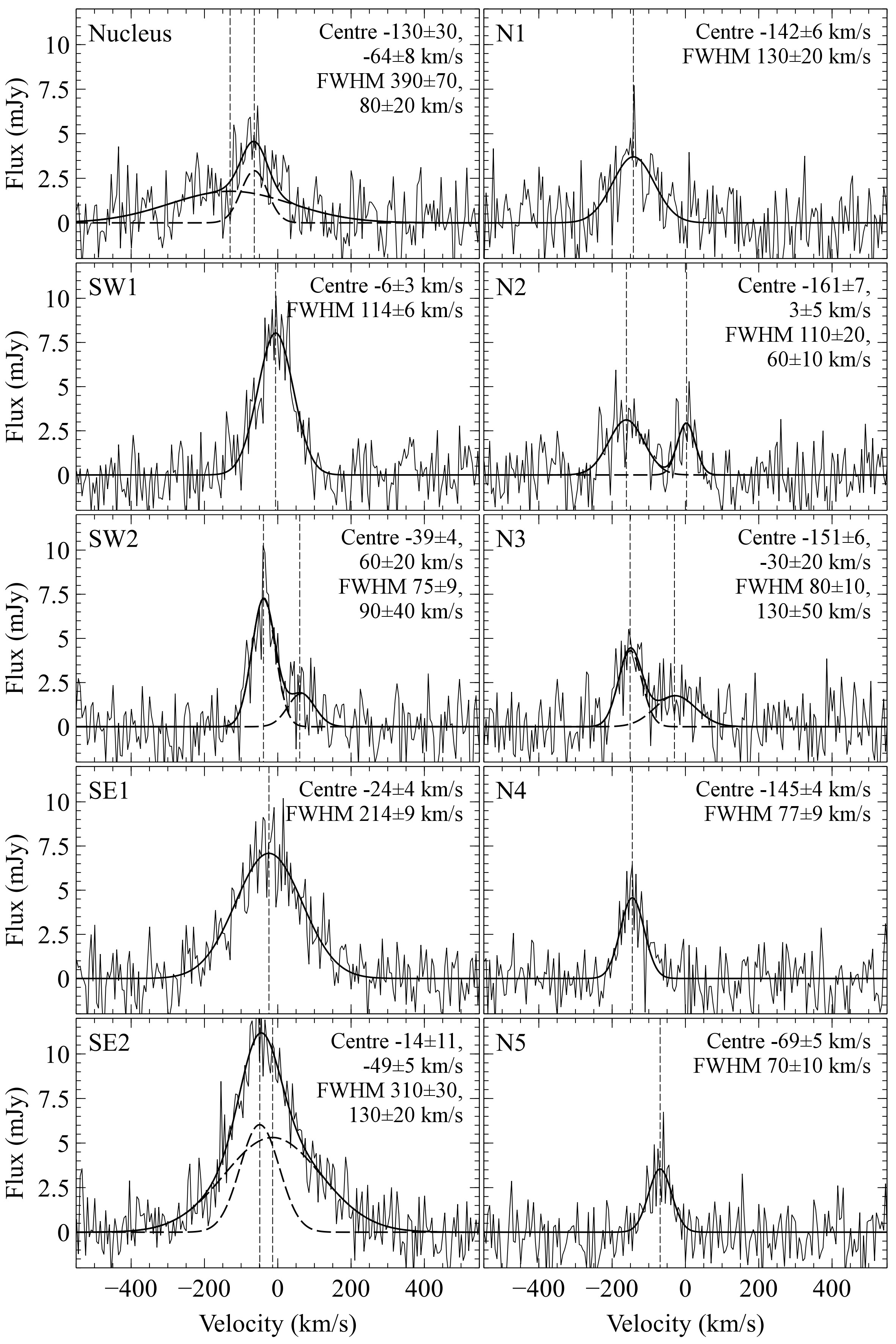}
\caption{CO(3-2) spectra for beam-sized regions along each filament with low numbers at small radii.  The line centres are marked with vertical dashed lines for comparison.}
\label{fig:CO32compspec}
\end{minipage}
\end{figure*}

Fig. \ref{fig:CO10vmaps} and Fig. \ref{fig:CO32vmaps} show the line
centre and FWHM for each of the detected velocity components at
CO(1-0) and CO(3-2), respectively.  The gas velocities lie within
$\pm100\kmps$ of the systemic velocity.  Furthermore the FWHM lies
below $200\kmps$ across most of the extended structure.  At CO(1-0),
the gas velocity centre ranges from $-87\pm6\kmps$ north of the
nucleus to $-50\pm10\kmps$ in the SE and $-27\pm4\kmps$ in the SW.
Additional velocity components are detected to the N and SE of the
nucleus.  These additional velocity components appear narrower than
the main component with FWHM of $40\pm20\kmps$ and $50\pm10\kmps$ to
the N and SE respectively.  Fig. \ref{fig:COtotalintensitywregs} shows regions selected to
cover these detections of multiple velocity components and
Figs. \ref{fig:CO102compspec} and \ref{fig:CO32compspec} show the spectra extracted from these regions
and the best-fit models.  Fig. \ref{fig:CO102compspec} shows the
best-fit models for the multiple velocity components detected in the N
and SE regions and for the SW region where a second component is not
significantly detected.  The second velocity component is redshifted
to $28\pm6\kmps$ to the N and blueshifted to $-212\pm5\kmps$ to the
SE.  In each case, the flux of the additional component is roughly
$10\%$ of the primary velocity component.

The velocity structure at CO(3-2) is clearly resolved into three main
filaments but appears broadly similar to the CO(1-0) structure.  The N
filament contains predominantly blueshifted gas with velocity centres
from $-161\pm7\kmps$ to $-142\pm6\kmps$ and this decreases to
$-49\pm5\kmps$ and $-39\pm4\kmps$ in the SE and SW filaments
respectively.  The FWHM is less than $150\kmps$ along the N and SW
filaments except for a region of redshifted emission to the NW where
the FWHM increases to $260\pm30\kmps$.  The N filament appears to bend
to the NW at its furthest extent and may be breaking up in this
region.  Additional velocity components are clearly detected in the N
and SE filaments, roughly coincident with the additional velocity
components detected at CO(1-0).  This second component has a velocity
centre of $-30\pm20\kmps$ to $3\pm5\kmps$ in the N filament and
$-49\kmps$ in the SE filament.  The FWHM is greatest at $200-300\kmps$
for the majority of the gas in the SE filament.  The additional
velocity components appear broader at CO(3-2) with FWHM of
$60-130\kmps$ in the N filament and $130\pm20\kmps$ in the SE
filament.  At CO(3-2), a second velocity component is also detected at
a radius of $1.6\asec$ ($3\kpc$) along the SW filament and a broad
component with FWHM of $390\pm70\kmps$ is coincident with the nucleus
(Figs. \ref{fig:CO32vmaps} and \ref{fig:CO32compspec}).  The fraction
of the flux in the second velocity component compared to the total
emission in each filament is $\sim10\%$, which is similar to the
fraction at CO(1-0).

In summary, the velocities of the molecular gas are low, lying within
$\pm100\kmps$ of the galaxy's systemic velocity.  The FWHM lies below
$200\kmps$ across most of the extended structure, which is
significantly below the stellar velocity dispersion of BCGs (eg. \citealt{vonderLinden07}).  The velocity
structure is therefore inconsistent with gravitational motions within
the galaxy.


\subsection{Velocity profiles}

\begin{figure*}
\begin{minipage}{\textwidth}
\centering
\hspace{-1cm}\includegraphics[width=0.48\columnwidth]{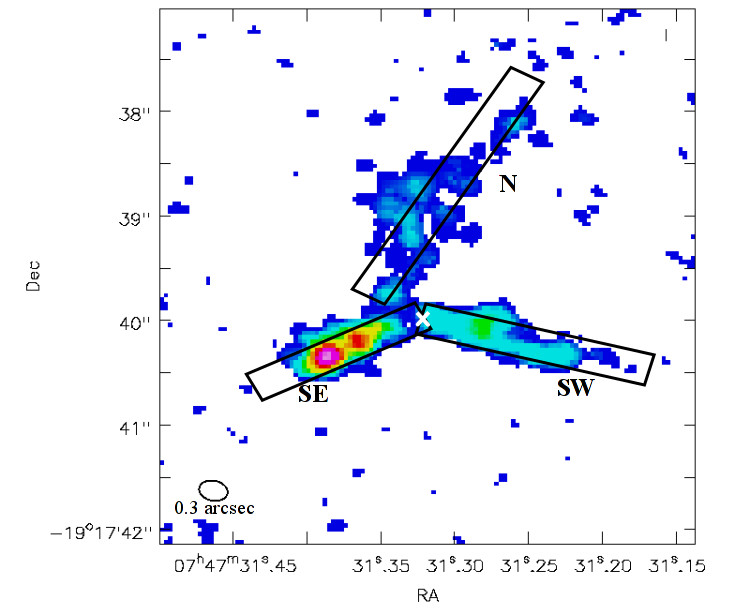}
\includegraphics[width=0.4\columnwidth]{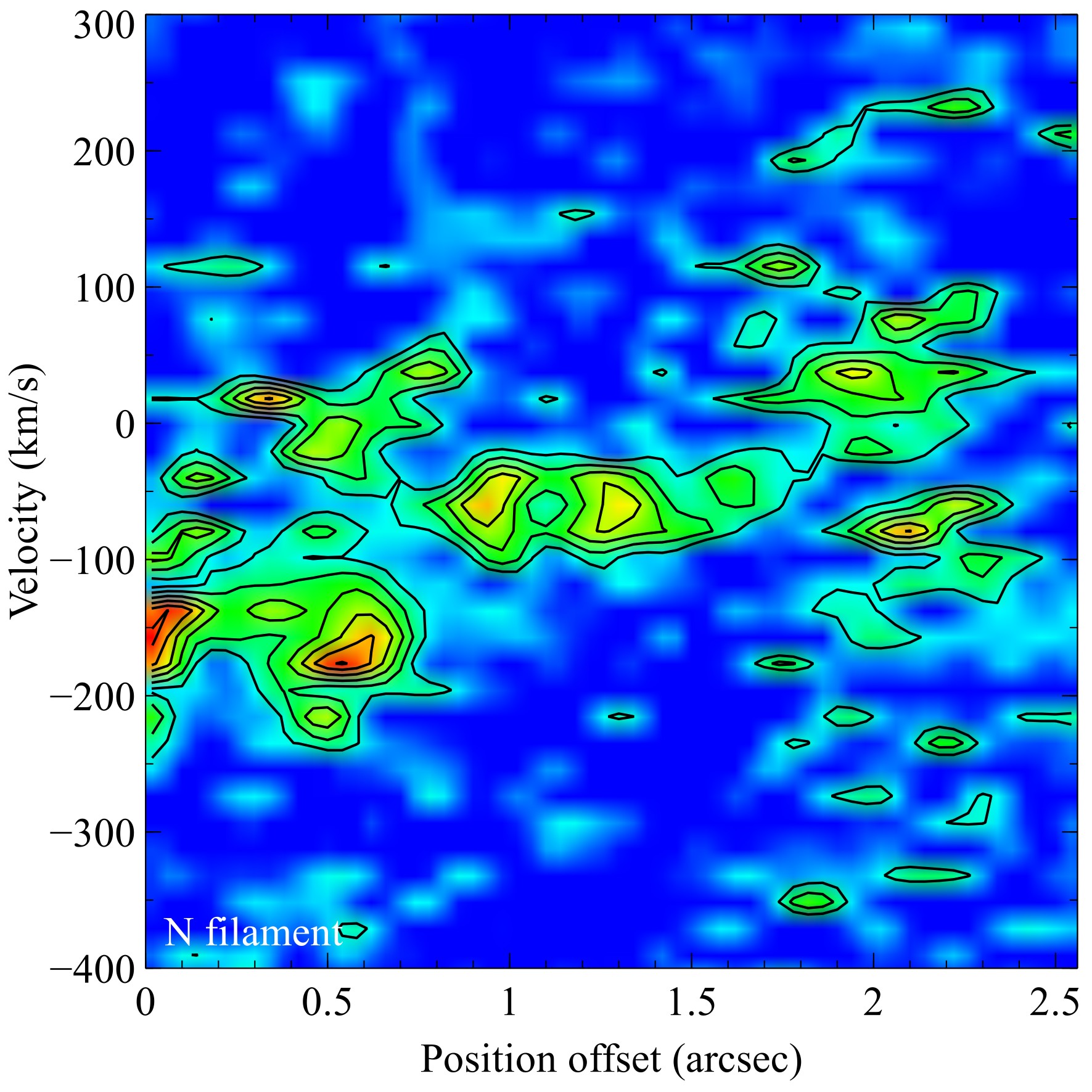}
\includegraphics[width=0.4\columnwidth]{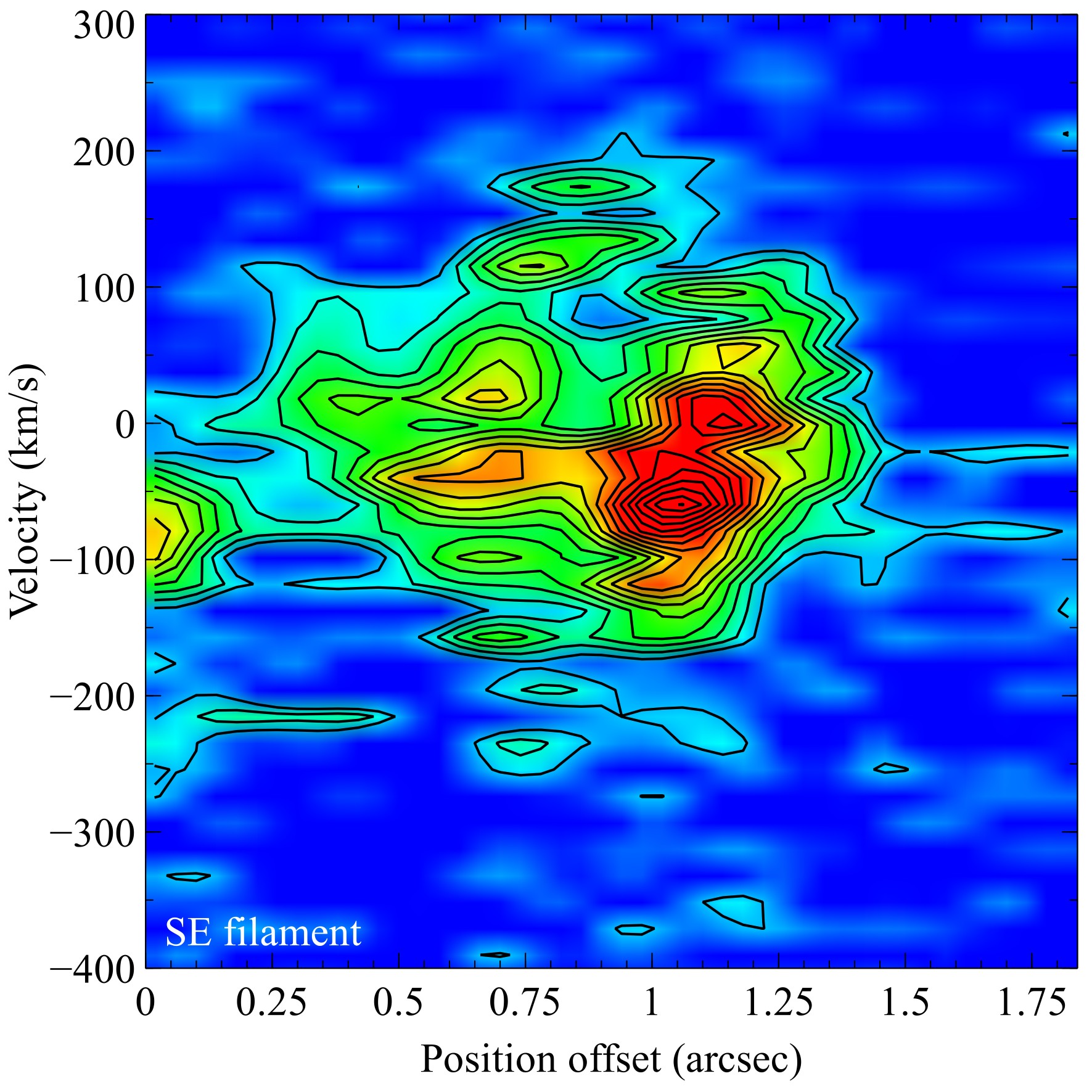}
\includegraphics[width=0.4\columnwidth]{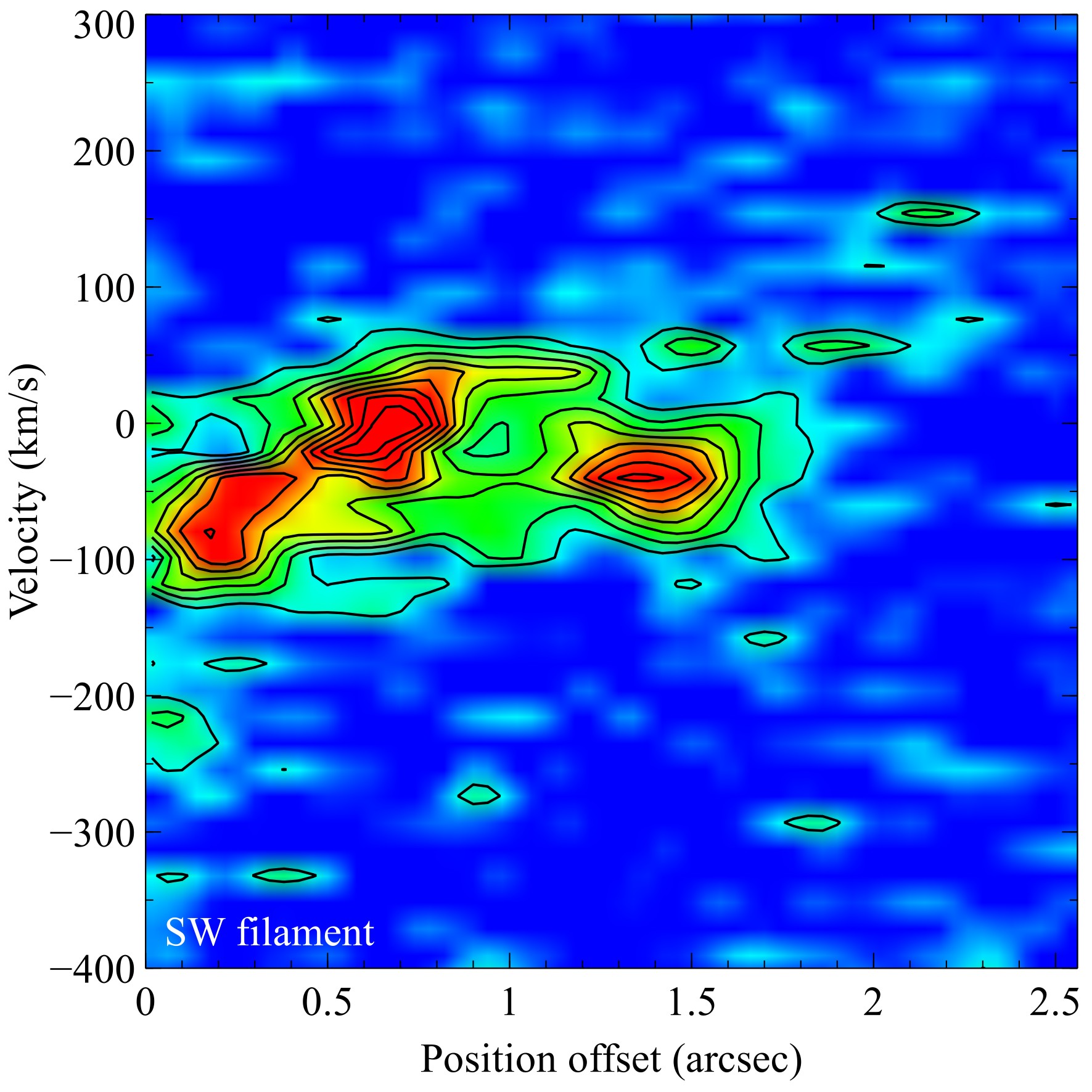}
\caption{Position velocity diagrams for each of the CO(3-2) filaments
  with the CO(3-2) integrated intensity map overlaid with the location
  of each PV slit shown upper left.  The position offset is minimum
  closest to the nucleus for each diagram.  Contours are $2\sigma$,
  $3\sigma$, $4\sigma$ etc.  The nuclear continuum point source location is shown by the cross (section \ref{sec:reduction}).}
\label{fig:CO32pv}
\end{minipage}
\end{figure*}

Fig. \ref{fig:CO32vmaps} shows ordered velocities along the N and SW
filaments.  We have produced position-velocity (PV) cuts along each of
the filaments resolved in the higher spatial resolution CO(3-2)
observations to identify velocity gradients (Fig. \ref{fig:CO32pv}).
Fig. \ref{fig:CO32pv} (upper left) shows the axis used for each PV
slice, which summed the line emission across the width of each
filament ($0.2-0.3\asec$).  The N filament has a shallow velocity
gradient from $-150\kmps$ to $50\kmps$ over $2.5\asec$ ($4.8\kpc$).
The inner region of this filament shows the two velocity components
detected within $1\asec$ of the nucleus at $-150\kmps$ and at the
BCG's systemic velocity.  The SE filament contains the broadest
velocity component with a FWHM of $300\kmps$.  The gas clouds in this
filament could be moving nearly along the line of sight and therefore
at a different orientation the velocity structure would be similar to
the other two filaments.  The two intensity peaks in the SE filament
have similarly broad FWHM and they may be physically separate along
the line of sight.


A velocity gradient from $50\kmps$ at large radii to $-100\kmps$
coincident with the nucleus is observed along the inner $1\asec$
($1.9\kpc$) of the SW filament.  This gradient may reverse at larger
radii or the outer part of the SW filament may be breaking up or a
separate structure.  The velocity structure of the SW filament appears
to link continuously to the gas projected across the nucleus with a
peak at $-64\pm8\kmps$ and this structure may extend to include the
$-150\kmps$ gas at the base of the N filament.  The SE and N filaments
appear clumpy towards the galaxy centre with no clear velocity
gradient across the nucleus.  However, the additional broad velocity
component at the centre suggests there could be a superposition of
fainter structures at a range of velocities.  The CO(1-0) observations
are consistent with this velocity structure but at lower spatial
resolution (Fig. \ref{fig:CO10vmaps}).




\subsection{Molecular gas mass}
\label{sec:mass}

The molecular gas mass can be calculated from the integrated CO(1-0)
intensity ($S_{\mathrm{CO}}\Delta\nu$) under the assumption of a
Galactic CO-to-H$_2$ conversion factor ($X_{\mathrm{CO}}$):

\begin{equation}
M_{\mathrm{mol}}=1.05\times10^4X_{\mathrm{CO}}\left(\frac{1}{1+z}\right)\left(\frac{S_{\mathrm{CO}}\Delta\nu}{\Jykmps}\right)\left(\frac{D_{\mathrm{L}}}{\Mpc}\right)^2\Msun,
\end{equation}

\noindent where $X_{\mathrm{CO}}=2\times10^{20}\COtoH$ in the Milky
Way disk (\citealt{Solomon87}; \citealt{Solomon05};
\citealt{Bolatto13}), $D_{\mathrm{L}}$ is the luminosity distance, and
$z$ is the redshift of the BCG.  However, the Galactic conversion
factor is not expected or observed to be universal
(eg. \citealt{Narayanan11}; \citealt{Bolatto13}), and likely depends
on many environmental factors such as the gas metallicity
(eg. \citealt{Wilson95}; \citealt{Arimoto96}; \citealt{Bolatto08}).
\textit{Chandra} observations show that the cluster atmosphere on the
scales of the BCG, from which the molecular gas likely cooled, has
subsolar metallicity of $0.4\Zsun$ with only modest spatial variations
(\citealt{Sanders14}).  Low metal abundance likely results in an
underestimate of the molecular gas mass in PKS\,0745-191, unless the cool
gas in the filaments has a higher metallicity compared to the ambient
ICM (eg. \citealt{Panagoulia13}).  

However, \textit{Spitzer} observations measure an IR luminosity for
the BCG just below the $10^{11}\Lsun$ threshold for a LIRG and
starburst galaxies are known to have $X_{\mathrm{CO}}$ factors lower
than found for galaxy disks.  Although the BCG in PKS\,0745-191 falls
short of the massive starbursts found in ULIRGs, the molecular gas
mass could be overestimated by a factor of a few (\citealt{Solomon97};
\citealt{Downes98}).  For kinematically disturbed and turbulent gas
associated with outflows where the gas may be optically thin even
lower $X_{\mathrm{CO}}$ factors are expected (for a review see
\citealt{Bolatto13}).  Although the molecular gas in PKS\,0745-191 is not
settled in the gravitational potential well, the gas velocities are
low.  The velocity dispersions of the molecular clouds are narrow
$<6\kmps$ (\citealt{David14}; Tremblay et al. in prep.), which is
typical of an individual molecular cloud in the Galactic disk.  In
summary, none of these factors suggest that the $X_{\mathrm{CO}}$
factor for the BCG in PKS\,0745-191 is anomalous and the low metal
abundance suggests the molecular gas mass could be underestimated.
Our conclusions are not qualitatively altered by the estimated factor
of a few uncertainty.


From the integrated CO(1-0) intensity of $2.2\pm0.1\Jykmps$ (Table
\ref{tab:fits}), the total molecular gas mass is
$(4.6\pm0.3)\times10^9\Msun$.  Roughly one third of this mass is
unresolved within the central peak close to the nuclear continuum
emission.  Assuming the CO(1-0) and CO(3-2) emission are distributed
similarly (section \ref{sec:lineratios}), we can estimate the molecular gas mass of each filament.
The SE filament is likely the most massive with
$\sim1.9\times10^{9}\Msun$.  The N and SW filaments have similar total
masses with $1.5\times10^9\Msun$ and $1.3\times10^9\Msun$,
respectively.  Although the CO(3-2) emission appears more centrally
concentrated with $\sim45\%$ of the total flux within a radius of
$0.8\asec$ (comparable to the CO(1-0) synthesized beam), this may
indicate that some extended structure has been resolved away rather
than intrinsic differences in the spatial distribution.

\subsection{Star formation in the BCG}
\label{sec:starformation}

Assuming that CO(3-2) emission traces the gas surface area, the
molecular gas surface density of $\sim800\Msunpsqpc$ for the BCG is
comparable to that of circumnuclear starburst galaxies
(eg. \citealt{Kennicutt98}; \citealt{Daddi10}; \citealt{Kennicutt12}).
HST FUV observations show a clear UV excess at the centre of the BCG
extending to the W of the nucleus that is clearly coincident with the
inner part of the SW filament before it bends at $\sim1\asec$ radius
(Fig. \ref{fig:hstvla}).  The SBC F140LP observation does not suffer
from contaminating line emission or a significant AGN contribution
(\citealt{Quillen08}; \citealt{Donahue11}; \citealt{Mittal15}).  The
cold molecular gas filament extending to the SW is therefore
coincident with a recent burst of star formation whilst the other two
filaments lie along dust lanes with little star formation detected in
the FUV (Fig. \ref{fig:hstvla}).  The dust lanes may obscure star
formation in the N and SE molecular filaments but the agreement
between the FUV and IR-derived star formation rates suggests no significant population of buried young stars
(\citealt{Johnstone87}; \citealt{Romanishin87};
\citealt{HicksMushotzky05}; \citealt{Donahue11}; \citealt{ODea08};
\citealt{Mittal15}).  \textit{Spitzer} infrared photometry and spectroscopy reveal star formation rates of
$17\Msunpyr$ (\citealt{ODea08}) and $11\Msunpyr$
(\citealt{Donahue11}), respectively.  Using the extent of the CO(3-2)
emission in the SW filament, the star formation surface density is
$\sim6\Msunpyrpsqkpc$.  The BCG in PKS\,0745-191 therefore lies on the
Kennicutt-Schmidt relation with IR-selected starburst galaxies
(\citealt{Kennicutt98}; \citealt{Kennicutt12}).


\section{Discussion}



\subsection{Origin of the molecular gas}
\label{sec:origin}

The inferred molecular gas mass in the BCG is substantially higher
than that typically found in early-type galaxies raising the question
of its origin.  Mergers between the BCG and donor galaxies are likely
rare due to the dearth of gas-rich galaxies at the centres of rich
clusters (eg. \citealt{Young11}) and because the cluster's high
velocity dispersion decreases the merger rate.  Ram pressure stripping
by the ICM of a gas rich galaxy that passes close to the BCG seems
similarly unlikely.  Molecular gas is much more tightly bound to its
host galaxy than atomic gas and is retained on all but the most
eccentric orbits within a cluster (eg. \citealt{Young11}).
Observations of Virgo cluster members have shown that galaxies with
strong HI deficiencies have minimal molecular gas deficiencies
(\citealt{Kenney89}; \citealt{diSerego07}; \citealt{Grossi09}).  In
PKS\,0745-191, the filaments have low velocities, radial morphologies and
shallow velocity gradients with no indication of the high velocities
and large-scale rotation expected from the stripping of a merging
galaxy (eg. \citealt{Ueda14}).  The three observed filaments in the
BCG would likely require separate direct impacts by gas-rich galaxies
each depositing similarly large quantities of molecular gas at low
velocities.

BCGs located at the centres of dense cluster atmospheres with
radiative cooling times below a Gyr are known to host cold molecular
gas in excess of a few $\times10^9\Msun$ (\citealt{Edge01};
\citealt{Salome03}).  The Perseus cluster hosts $4\times10^{10}\Msun$
of extended molecular gas filaments coincident with the coolest X-ray
gas (\citealt{Salome06}).  The simplest interpretation of their
velocity structure is an inflow of gas cooling from the ICM,
free-falling toward the BCG (\citealt{Salome08}; \citealt{Lim08}).  In
PKS\,0745-191, the X-ray cooling rate of $270\pm90\Msunpyr$ determined
from XMM-RGS spectroscopy (\citealt{Sanders14}) could supply the
molecular gas mass in only $\sim20\Myr$.  This cooling timescale is
also comparable to the time since the last major AGN outburst given by
the age of the X-ray cavities, which require $\sim10-20\Myr$ to rise
at the sound speed to the observed projected radii
(\citealt{Rafferty06}; \citealt{Sanders14}).  Whilst it appears likely
that the molecular gas cooled from the cluster atmosphere, the low gas
velocities are inconsistent with free-fall.



\subsection{Molecular gas clouds in gravitational free-fall?}


The velocity gradients detected in the N and SW filaments could
indicate an inflow of gas cooling from the cluster atmosphere but the
interpretation depends on the location of the gas along
the line of sight.  Without absorption line observations showing
whether the cold gas lies in front or behind the galaxy, it is
difficult to unambiguously determine whether gas is moving in or out.  However, the low velocities and shallow velocity gradients are inconsistent
with simple models for gas cooling steadily from the cluster atmosphere.  The
ICM is approximately in hydrostatic equilibrium; therefore, an
overdense, rapidly cooling gas blob is initially at rest with respect
to the cluster.  This cooling gas decouples from the cluster
atmosphere and is expected to subsequently free-fall in the cluster's gravitational
potential (eg. \citealt{PizzolatoSoker05,Pizzolato10};
\citealt{Gaspari15}).  Although X-ray gas pressure and cloud-cloud
collisions will slow the infall, in general we expect a smooth radial velocity gradient
with the highest velocities at the smallest radii.  The velocity
gradients along the SW and N filaments steadily increase to
$-100\kmps$ and $-200\kmps$ with decreasing radius over a distance of
a few kpc, respectively.  The SE filament may have a similar
velocity gradient obscured by an orientation close to the line
of sight.

\begin{figure*}
\begin{minipage}{\textwidth}
\centering
\includegraphics[width=0.32\columnwidth]{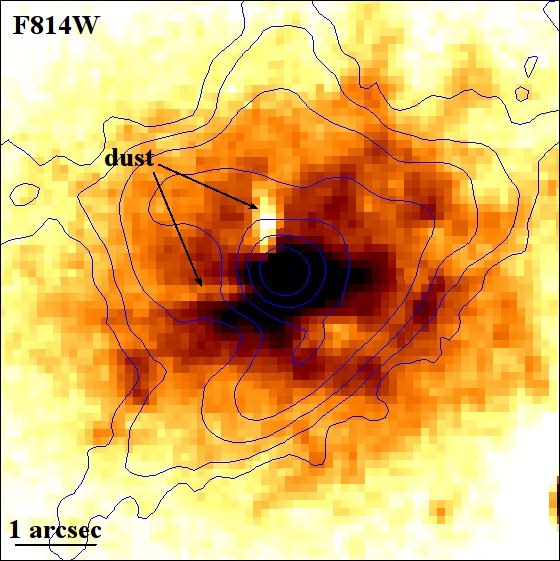}
\includegraphics[width=0.32\columnwidth]{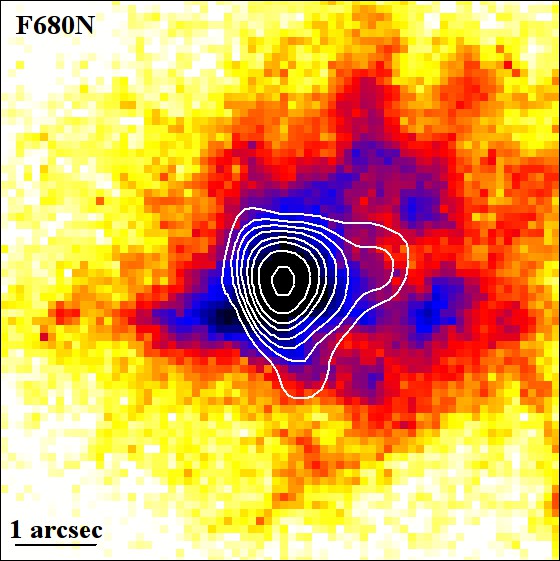}
\includegraphics[width=0.32\columnwidth]{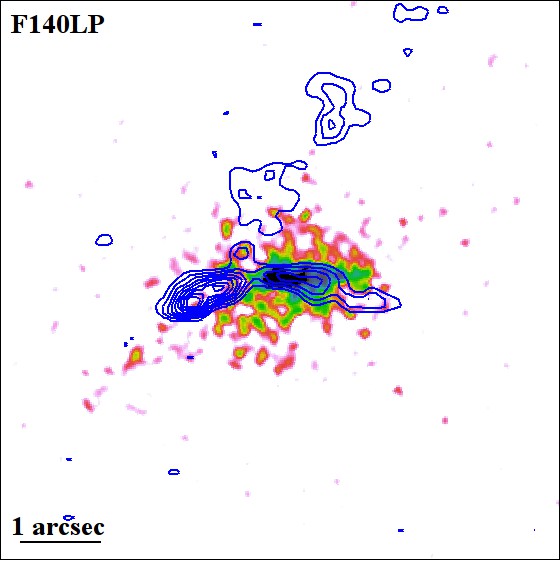}
\caption{Left: HST F814W image of the brightest cluster galaxy showing the old stellar population and H$\alpha{+}$[N\textsc{ii}] emission (\citealt{Sand05}; \citealt{Mittal15}; \citealt{Tremblay15}).  A smooth model has been subtracted to show the dust lanes extending to the N and E of the nucleus.  Contours show the $5\GHz$ VLA radio emission from \citet{Baum91}.  Centre: Narrow band image covering the H$\alpha{+}$[N\textsc{ii}] emission (\citealt{Donahue00}) with the $14.9\GHz$ VLA B array contours overlaid (\citealt{Baum91}).  Right: F140LP SBC image of FUV continuum emission from massive, young stars (\citealt{Mittal15}; \citealt{Tremblay15}) with ALMA CO(3-2) integrated intensity contours overlaid.  All images are centred on the nuclear continuum point source emission (section \ref{sec:reduction}).}
\label{fig:hstvla}
\end{minipage}
\end{figure*}


Following \citet{Lim08}, we assume a Hernquist model for the
gravitational potential of an elliptical galaxy, expressed in terms of
the total mass and effective radius, that reproduces a de Vaucouleurs
surface brightness profile (\citealt{Hernquist90}).  The velocity of a
gas blob that free-falls in this potential is given by

\begin{equation}
v(r)^2=v(r_0)^2 + 2GM\left(\frac{1}{r+a} - \frac{1}{r_0+a}\right),
\end{equation}

\noindent where $M$ is the total galaxy mass, $a$ is the scale radius,
which is related to the effective radius $R_e\sim1.8153a$, $r_0$ is
the initial radius and $v(r_0)$ is the initial velocity.  For
PKS\,0745-191, \citet{Allen96} measured a projected mass within a radius
of $34\kpc$ of $1.8\times10^{13}\Msun$ using a strong gravitational
lensing analysis (corrected for our cosmology, see also
\citealt{Sanders14}).  However, X-ray mass estimates assuming
hydrostatic equilibrium have measured a lower total mass of
$\sim2\times10^{12}\Msun$ within a radius of $30\kpc$
(\citealt{Voigt06}; \citealt{Sanders14}).  The X-ray mass may be
affected by strong deviations from spherical symmetry or a breakdown
of hydrostatic equilibrium in the cluster centre.  We have considered
both estimates of the total mass in our analysis.  The scale radius of
$10\kpc$ was estimated from the K-band effective radius of $9.5\asec$
given by the 2MASS extended source
catalogue\footnote{http://www.ipac.caltech.edu/2mass}
(\citealt{Skrutskie06}).  We note that this may underestimate the true
value as the 2MASS observations are not deep enough to trace the full
extent of the BCG envelope and PKS\,0745-191 lies near the Galactic plane
on the sky.

The remaining free parameters are the initial radius, where the gas
blob starts to free-fall, and the inclination to the line of sight.
Beyond $\sim0.5\arcsec$ ($\sim1\kpc$) from the initial radius, the
velocity increases linearly with decreasing radius therefore the
inclination is the main parameter that can be altered to match the
velocity structure.  However, for both the N and SW filaments, the
shallow velocity gradients require an orientation $<20^{\circ}$ from
the plane of the sky.  This assumes the conservative X-ray estimate of
the total mass.  The gravitational lensing total mass requires an even
more stringent orientation $<10^{\circ}$ from the plane of the sky.

The low velocities of the cold gas found in all three filaments appear
inconsistent with gravitational infall.  For an orientation of
$30^{\circ}$ from the plane of the sky, a gas blob released from rest
at a projected radius of $2\asec$ ($3.8\kpc$) will reach a velocity of
$\sim250\kmps$ over $1\asec$ projected distance.  Depending on the
exact form of the gravitational potential, the expected free-fall velocities
could be higher by a factor of two.  Therefore, if the molecular gas
structures originated more than a few kpc from their current locations
their velocities should significantly exceed those observed.  The
velocity of an infalling gas blob may be slowed by ram pressure drag
from the ICM, although this is expected to be a minor effect for such
dense gas, or by cloud-cloud collisions inside a radius of $\sim1\kpc$
(eg. \citealt{PizzolatoSoker05}; \citealt{Gaspari15}).  Cloud-cloud
collisions could be producing the broader FWHM at the base of the N
filament where it may intersect the other filaments but are unlikely
to substantially reduce the gas velocities at larger radii.  For such
low velocities and narrow FWHM, the molecular filaments must be
transient structures formed from gas cooling locally and supported in
situ (see section \ref{sec:discexfils}).


If the molecular filaments formed from rapid cooling of the hot
atmosphere, the lack of molecular gas that could be associated with a
previous cooling episode appears inconsistent with the observed
stability of AGN feedback.  The prevalence of short central radiative
cooling times in cluster atmospheres and strong correlations with
detections of cold gas, star formation and H$\alpha$ emission suggest
that cooling is long-lived (\citealt{McNamara12};
\citealt{Fabian12}).  Studies of the fraction of cool core clusters
with radio bubbles and central radio sources suggest that the duty
cycle of AGN activity in BCGs is at least $70\%$
(\citealt{DunnFabian06,DunnFabian08}; \citealt{Birzan12}).  AGN
feedback appears more or less continuous, rather than strongly
episodic, and the good agreement between the radio power and cooling
losses suggest a stable balance to $z\sim0.7$ (eg. \citealt{Ma11};
\citealt{HlavacekLarrondo12}).  Repeated episodes of cooling
and heating in PKS\,0745-191 are more consistent with these observations
than a single, sudden influx of cold gas onto the BCG.  Rapid
accretion of previous massive molecular inflows by the SMBH would require
implausibly high efficiency forming few young stars in the process.

\subsection{Outflowing molecular gas clouds?}


The most significant dust lane features detected in the BCG in HST
observations (Fig. \ref{fig:hstvla}; \citealt{Donahue00}) lie along the
N and SE filaments.  If the dusty molecular gas clouds obscured more
than $\sim50\%$ of the galaxy light then blueshifted and redshifted
gas components could be cleanly interpreted as outflow and inflow,
respectively.  Using the F555W \textit{HST} image, we subtracted the
average surface brightness determined in sectors free of emission
lines and dust in a series of elliptical annuli.  The ellipticity and
position angle for the BCG were taken from the 2MASS extended source
catalog (\citealt{Skrutskie06}) and should therefore not be affected
by the line emission and dust lane structure.  Along the dust lanes
coincident with the N and SE filaments, the surface brightness falls
$20-50\%$ below the average at that radius.  The SW filament is
instead coincident with excess emission at $20-40\%$ above the average
at each radius likely due to star formation.

The peak absorbed fraction in the N and SE filaments could be higher
as they are likely composed of unresolved dense, giant molecular cloud
associations producing variations in the covering fraction on
unresolved spatial scales in the \textit{HST} images
(eg. \citealt{Salome08,Salome11}; \citealt{David14}).  In NGC\,1275,
\citealt{Salome08} found that the CO line ratios indicate optically
thick radiation but the observed brightness temperature of the
molecular gas clumps is at least an order of magnitude below the
expectation for normal, optically thick clouds.  These observations
can be reconciled if each molecular clump is comprised of many giant
molecular clouds each with a mass of $\sim10^6\Msun$ and a radius of
$\sim30\pc$ typical of those found in the Milky Way
(\citealt{Solomon87}).  Although the complexity of the structure makes
a clean interpretation difficult, it is plausible that the peak
absorbed fraction is significantly higher and therefore the bulk of
the molecular gas in the NW and SE filaments lies on the near side of
the galaxy.  Regions of $\sim50\%$ absorption in the N and SE
filaments show that some of the gas clouds lie in front of the galaxy
midplane.

Both velocity components in the SE filament are blueshifted with
respect to the BCG's systemic velocity and may be flowing away from
the galaxy centre.  Whilst the brightest velocity component at CO(1-0)
has only a small velocity shift to $-29\pm5\kmps$, the second
component is clearly observed as a blueshifted wing at
$-212\pm5\kmps$.  Although two velocity components are detected in the
SE filament in the CO(3-2) emission, the velocities are modest,
perhaps indicating different gas properties in this outflowing
component.  Blueshifted and redshifted velocity components are
detected in the N filament at both CO(1-0) and CO(3-2).  Multiple
coincident velocity components likely indicates a superposition of
structures along the line of sight.  The more massive blueshifted
component presumably lies in front of the galaxy centre and may
similarly indicate gas outflowing from the galaxy centre.  The second,
less massive velocity component is redshifted to at most $28\pm6\kmps$
and is projected close to the nucleus but may lie anywhere along the
line of sight.

In summary, the velocity structure and dust obscuration observed in
the N and SE filaments are consistent with the bulk of the molecular
gas moving outwards from the galaxy centre.

\subsection{Molecular gas clouds lifted by radio bubbles?}

\begin{figure}
\centering
\includegraphics[width=0.95\columnwidth]{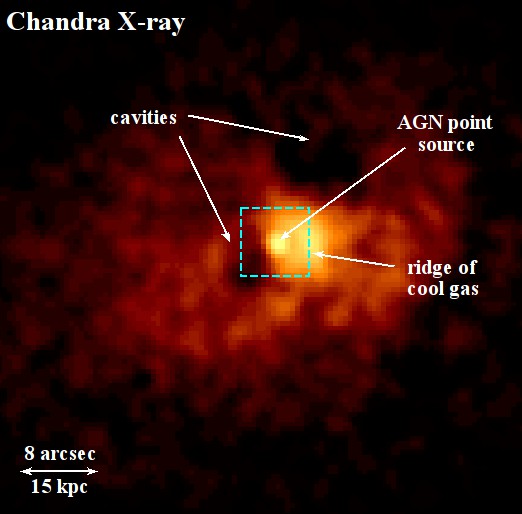}
\caption{\textit{Chandra} X-ray image ($0.3-7\keV$) of the hot
  atmosphere surrounding the BCG in PKS\,0745-191.  A smooth X-ray
  background has been subtracted.  The box corresponds to the region
  covered by the optical images in Fig. \ref{fig:hstvla}.}
\label{fig:xray}
\end{figure}

Radio and X-ray observations suggest that radio jet outbursts from the
central SMBH are interacting with the extended molecular and ionized
gas filaments.  The VLA $14.9\GHz$ observation with a resolution of
$\sim0.15\arcsec$ shows a bright nucleus with an extension of the
radio emission $0.5\arcsec$ in length to the W (Fig. \ref{fig:hstvla};
\citealt{Baum91}).  The bright radio nucleus is coincident with the
continuum point source detected with ALMA in the sub-mm and a hard
X-ray point source detected with \textit{Chandra}.  A jet may have
disrupted in the dense cluster environment, which is consistent with
the amorphous radio appearance.  Other radio galaxies in dense cluster
environments are similarly disrupted, such as PKS\,1246-410 at the
centre of the Centaurus cluster (\citealt{Taylor06}). This more
diffuse structure observed in PKS\,0745-191 in $1-2\GHz$ VLA images
extends towards two X-ray surface brightness depressions, each
$\sim20\kpc$ across, located to the N and SE in \textit{Chandra}
observations (Fig. \ref{fig:xray}; \citealt{Rafferty06};
\citealt{Sanders14}).  The AGN has inflated two large radio bubbles
detected as cavities in the X-ray emission where the radio lobes have
displaced the cluster's hot atmosphere (\citealt{McNamaraNulsen07};
\citealt{Fabian12}; \citealt{McNamara12}).

The N and SE molecular filaments extend out towards the X-ray cavities
and may have been lifted from the galaxy in the wake of the
buoyantly rising bubbles.  No clear cavity is detected near the SW
molecular gas filament although this would be difficult to detect
under the bright soft X-ray ridge of emission.  A small, young cavity
may be located between the ridge and the X-ray point source where
the X-ray surface brightness drops.  ALMA observations of the BCG at
the centre of A1835 have found a $10^{10}\Msun$ bipolar molecular flow in similarly extended filaments that may be accelerated outward by the mechanical energy associated with rising radio bubbles (\citealt{McNamara14}).
Single dish detections of molecular gas coincident with H$\alpha$
filaments beneath buoyant radio bubbles in the Perseus cluster are
consistent with this scenario (\citealt{Salome06,Salome11}).  Radio
jet-driven outflows of molecular gas have also been detected in nearby
galaxies (eg. \citealt{Alatalo11}; \citealt{Tadhunter14};
\citealt{Morganti15}).  Using a total molecular gas mass of
$5\times10^9\Msun$, a gas velocity of $100\kmps$ and velocity
dispersion of $100\kmps$, the energy required to accelerate the molecular gas is
several $\times10^{57}\erg$.  By calculating the $4PV$ work done
inflating the two radio bubbles in the hot atmosphere,
\citet{Sanders14} estimated that the AGN outburst supplies
$\sim3\times10^{60}\erg$, which is energetically sufficient to uplift the
molecular gas.  Buoyant bubbles cannot lift more weight than they
displace.  Using a conservative estimate of the surrounding gas
density, the radio bubbles displace $\sim5\times10^{10}\Msun$ of hot
gas and therefore could have lifted $5\times10^9\Msun$ of molecular
gas along with a significant amount of warmer gas.

The NW and SE filaments in PKS\,0745-191 comprise $60-70$ per cent of the
total cold molecular gas, which suggests very strong coupling to the
radio bubbles.  The lack of any centrally concentrated reservoir of
molecular gas coincident with the nucleus requires an even higher
coupling fraction.  It is difficult to understand how radio bubbles
could uplift dense molecular gas clouds so efficiently.
Volume-filling gas would be easier to lift.  Roughly 20 per cent of
the molecular gas appears uplifted by the radio bubbles in A1835,
which together with the hot gas outflow produces a total uplifted mass
uncomfortably close to the theoretical maximum.  \citet{McNamara14}
suggested that turbulence maintained by ongoing star formation would
reduce the density contrast and enable uplift of the molecular gas by
the hotter gas.  However, the star formation in PKS\,0745-191 is
concentrated in the SW filament.  


\subsection{Forming molecular gas in extended filaments}
\label{sec:discexfils}

In a cluster atmosphere where AGN heating locally balances radiative
cooling, uplift by radio bubbles can promote cooling by removing gas
from this equilibrium state.  Low entropy gas that has been lifted to
larger radii is cooler and denser than the surroundings and therefore
radiates more energy than can be replaced by the local heating rate.
The molecular gas filaments could therefore have formed from warmer
gas that was uplifted by the radio bubble and cooled locally
(\citealt{Churazov01}; \citealt{Revaz08}; \citealt{Li14}).  Rapid cooling of shock-heated outflowing gas to form molecules on short timescales has been suggested to explain the molecular gas kinematics in IC\,5063 and 4C\,12.50 (\citealt{Morganti13}; \citealt{Tadhunter14}; \citealt{Zubovas14}; \citealt{Morganti15}). Fig. \ref{fig:hstvla} shows
that the molecular gas filaments also trace the brightest regions of
optical line emission and this gas is coincident with low temperature
X-ray gas (\citealt{Sanders14}).  The time taken for the radio bubbles
to rise at the sound speed to the observed projected radii is
$\sim10\Myr$ for the E bubble and $\sim20\Myr$ for the NW bubble.  The
X-ray cooling rate of $270\pm90\Msunpyr$ is large enough to supply the
inferred molecular gas mass of the filaments in this time (section
\ref{sec:origin}).  This would however require catastropic cooling of
the hot gas around the radio bubbles.  From \citet{Sanders14}, the hot
gas density profile within $\sim20\kpc$ radius can be approximated by
a powerlaw $\rho\propto{r^{-1}}$.  Extrapolating into the cluster
centre, the hot gas mass is $\sim4\times10^9\Msun$ within a sphere of
radius $5\kpc$, which covers the extent of the molecular filaments.
Rapid cooling to supply the $\sim5\times10^9\Msun$ of molecular gas
would therefore deplete the hot gas at the cluster centre and the
resulting inflow would oppose uplift of the gas.


The slowest step in the cooling process is the formation of molecular
hydrogen, which requires the presence of dust grains to occur on these
timescales (eg. \citealt{Ferland94,Ferland09}).  Dust grains are
sputtered rapidly ($<1\Myr$) in the X-ray atmosphere
(\citealt{Draine79}; \citealt{Dwek92}) so cooling X-ray gas is likely
dust-free.  However, the N and SE molecular filaments are clearly
coincident with dust lanes in the \textit{HST} observations
(Fig. \ref{fig:hstvla}).  Little recent star formation is detected in
these filaments but dust could have been distributed locally by the
BCG's older stellar population.  Material ejected from stars at the
centres of cool core clusters may not necessarily mix rapidly with and
be heated by the surrounding hot gas environment (\citealt{Voit11}).
If this gas remains cool the embedded dust will be preserved and could
enrich the cooling gas filaments.  The dusty molecular wake extending
$4\pc$ behind the evolved star Mira in our galaxy suggests that
stellar ejecta can remain cold and dense in the interstellar medium,
consistent with this scenario (\citealt{Martin07}).  Dust could also
have been uplifted from the BCG centre along with the rapidly cooling
gas and shielded from the X-ray atmosphere in dense gas clumps.  If
the molecular gas has the momentum of the warmer, uplifted gas that it
cooled from, the low outflow velocities suggest a gentle acceleration
that would not destroy dust grains.  Slow entrainment of the gas by
the radio bubbles would also not dissociate the molecular gas.


For the observed gas velocities of $\pm100\kmps$ and projected lengths
of $3-5\kpc$, the dynamical ages of the SE and N filaments are
$1-2\times10^7\yr$, which is approximately the bubble rise time.
However, if the molecular filaments are only supported by thermal
pressure, they would collapse under tidal gravitational forces on
timescales $<1\Myr$ (eg. \citealt{Ho09}).  The survival of dense gas filaments at the
centres of hot, high pressure cluster atmospheres for at least a
dynamical timescale suggests that magnetic fields are required to
support and insulate against rapid evaporation.  This assumes that the
small ionised fraction of the cold gas supported by the magnetic field
effectively transmits the magnetic stress to the large neutral portion
through collisions.  Ambipolar diffusion is also assumed to be
sufficiently slow to allow effective links between the molecular,
atomic and ionised gas components (eg. \citealt{Shu92};
\citealt{McKee07}).  \citet{Fabian08} show that the thread-like
H$\alpha$ filaments in the Perseus cluster are likely supported by
magnetic fields of at least a few tens of $\muG$ (see also
\citealt{Ho09}).  The filamentary structure of the molecular gas and
optical line emission in PKS\,0745-191 appears very similar to the Perseus
cluster.  The molecular gas filaments are likely to be a superposition
of many separate structures and unresolved strands.

However, the demands on magnetic support appear much greater in
PKS\,0745-191.  The magnetic pressure required to support the molecular
gas is $P_{\mathrm{B}} \sim B^{2}/8\pi \sim {\Sigma}g$, where $B$ is
the magnetic field strength, $\Sigma$ is the molecular gas surface
density and $g$ is the gravitational acceleration.  The width of each
filament in PKS\,0745-191 is not resolved therefore we calculate a lower
limit on the magnetic pressure by assuming that the width is given by
the synthesized beam.  The SE filament is $\sim3\kpc$ in length with a
radius of $0.3\kpc$ and a molecular gas mass of $1.9\times10^9\Msun$.
Following \citet{Fabian08}, for a radial filament the lengthwise
column density $N\sim7\times10^{23}\pcmsq$ and the surface density
$\Sigma_{\parallel}\sim1.2\gpcmsq$.  The gravitational acceleration at
the centre of PKS\,0745-191 $g\sim10^{-8}\cmpssq$ (\citealt{Sanders14}).
The magnetic pressure required
$P_{\mathrm{B}}\sim1.2\times10^{-8}\ergpcmcu$, which is roughly an
order of magnitude greater than the thermal pressure of the hot gas
$P_{\mathrm{T}}\sim8\times10^{-10}\ergpcmcu$ (\citealt{Sanders14}).
Therefore, supporting the weight of the massive cold gas filament
requires a magnetic pressure that substantially exceeds the hot gas
pressure.  The corresponding magnetic field strength
$B\sim400\muG$ is significantly above typical values for cluster cores
(\citealt{Carilli02}; \citealt{Govoni04}) and possible enhancements
from flux-freezing (eg. \citealt{Sharma10}).  Such a large, aspherical
pressure distribution in the cluster centre should produce departures
from hydrostatic equilibrium in the hot gas.  \citet{Sanders14} found
evidence for non-thermal pressure within a radius of $40\kpc$ in
PKS\,0745-191 and showed that the coolest X-ray emitting gas is offset by
around $5\kpc$ from the nucleus.  This asymmetry could be related to
sloshing of the cool core (\citealt{Sanders14}) or, for a
magnetically-dominated core, it could indicate clumping of the hot gas
as the ICM is squeezed by the magnetic pressure.  The hot gas in the
cluster centre will therefore have a low filling factor and a lower
mean density compared to a uniformly distributed medium.  The
amorphous structure of the central radio source is coincident with and
may be related to this magnetically-dominated core region.  Faraday
rotation measurements of the extended radio emission could be used to
determine the magnetic field strength.


The problems of magnetic pressure support and hot gas depletion could
be alleviated if the $X_{\mathrm{CO}}$ factor for BCGs, and
correspondingly the molecular gas mass, was lower by an order of
magnitude.  For a lower molecular gas mass of $\sim5\times10^8\Msun$,
the magnetic pressure required to support the filaments would be
comparable to the hot gas pressure, consistent with the X-ray
observations (\citealt{Sanders14}).  Rapid cooling of only 10\% of the
hot gas within a radius of $5\kpc$ could supply the molecular gas mass
on the required timescales.  A reduced $X_{\mathrm{CO}}$ factor would
also not move the filaments in PKS\,0745-191 significantly beyond the
scatter of the Kennicutt-Schmidt relation (eg. \citealt{Kennicutt12}).
ULIRG and starburst galaxy environments are known to have
kinematically disturbed molecular gas with higher temperatures and
velocity dispersions causing reductions in $X_{\mathrm{CO}}$ by a
factor of a few to ten (eg. \citealt{Solomon97}; \citealt{Downes98};
\citealt{Bolatto13}).  However, as discussed in section
\ref{sec:mass}, the low gas velocities in PKS\,0745-191 and narrow
velocity dispersions observed for molecular clouds in BCGs are
inconsistent with the conditions in starburst galaxies.  Herschel
observations tracing the dust emission in PKS\,0745-191 show that the bulk
of the dust is low temperature at $\sim25\K$ and a Galactic
$X_{\mathrm{CO}}$ factor produces a gas-to-dust mass ratio of
$\sim100$ consistent with that found in the Milky Way (Oonk et al. in
prep.).  Therefore, whilst a lower $X_{\mathrm{CO}}$ factor could
alleviate several problems, we find no clear evidence to support a
significantly lower value in PKS\,0745-191 beyond galaxy-to-galaxy
variations.  The low metal abundance suggests the molecular gas mass
could instead be underestimated unless the cool gas in the filaments has an
enhanced metallicity over the ambient ICM (eg. \citealt{Panagoulia13}).

\subsection{Fate of the molecular gas}

Fig. \ref{fig:hstvla} shows a striking coincidence between the SW
filament of molecular gas, a spur of young, massive stars in the FUV
and extended radio emission at detected at $7\sigma$ significance in
$14.9\GHz$ VLA observations (\citealt{Baum91}).  The filament also
extends towards the surface brightness peak of the cluster
emission.  In comparison, little recent star formation is observed
coincident with either the N or SE filaments even though each has a
higher inferred molecular gas mass and similar velocity structure.  An
external perturbation, such as an expanding radio structure, may have
triggered gas cloud collapse and star formation in the SW filament.
The dust lanes coincident with the N and SE filaments could obscure
recent star formation, although IR observations are inconsistent with
a substantial population of buried young stars (section
\ref{sec:starformation}) and it is not clear why the SW filament is
not similarly dusty.


A collision between an expanding radio lobe and dense gas clouds in
the SW filament may have disrupted the radio structure and triggered
star formation at this location, consistent with the amorphous radio
morphology in lower frequency observations.  Jet-triggered star
formation is thought to be occuring in Cen A and Minkowski's object
(\citealt{Graham81}; \citealt{vanBreugel85}; \citealt{Dey97};
\citealt{Bicknell00}; \citealt{Oosterloo05}; \citealt{Santoro15}; \citealt{Salome16}) and could explain the $\sim90^{\circ}$
deflection of the radio jets in A1795 (\citealt{McNamara96};
\citealt{Pinkney96}).  For A1795, knots of young star clusters and
dense molecular gas clouds clearly trace the outer edges of radio
lobes and an increase in the ionization state, turbulence and density
of the gas at the deflection point in the jets implies a direct
interaction (\citealt{Crawford05}). However, although the gas filament
appears deflected at the end of the small-scale extension of the radio emission in
PKS\,0745-191, no increase in the velocity dispersion is observed either
along the SW filament or by comparison with the other filaments.  The SW filament also has a lower CO(3-2)/CO(1-0) line ratio than the other filaments which is consistent with more diffuse gas.  It
is therefore not clear if these structures are physically interacting
or merely observed as coincident in projection.  


Star formation also occurs in a minority of the extended cool gas
filaments in the Perseus cluster despite their substantial molecular
gas mass (\citealt{Canning14}).  Only three separate regions of the
outer filaments have been disrupted by an unknown mechanism and
rapidly collapsed into stars $<50\Myr$ ago (\citealt{Canning14}).  The
filaments do not appear to become generally unstable due to bulk
sloshing motions in the ICM or growth through hot gas mixing with the
cold phase and weak shocks generated by the radio bubble inflation do
not appear to have similarly destabilised the inner filaments
(\citealt{Fabian11}; \citealt{Canning14}).  Star formation only in the
SW filament in PKS\,0745-191 is consistent with these findings and given
the comparable structure of the three separate gas filaments it
appears likely that the radio structure is responsible for triggering its
collapse.  

The long-term state of the non-star-forming filaments is unclear.
The gas velocities lie far below the BCG's escape velocity therefore
the molecular gas clouds will remain in the galaxy's potential.  The
outer filaments may eventually evaporate in the hot cluster gas or
fall back onto the central galaxy and subsequently fuel the AGN.  If
these cold gas clouds are subsequently fuelling the AGN, the feedback
must somehow be prompt or approximately constant to explain the close
connection between detections of cold gas, H$\alpha$ emission and star
formation and short radiative cooling times in the hot cluster
atmosphere (\citealt{Edge01}; \citealt{Salome03};
\citealt{Rafferty08}; \citealt{Cavagnolo08}).  Lower significance
structures at the outer sections of the N and SE filaments hint at the
break up of these filaments (Fig. \ref{fig:CO32vmaps}).  Deeper
observations will be required to determine the fate of these gas
clumps, potentially as a circulation flow falling back to the BCG
centre (\citealt{Lim08};
\citealt{Salome06,Salome11}).  





\section{Conclusions}

ALMA Cycle 1 observations of PKS\,0745-191 have revealed
$4.6\pm0.3\times10^9\Msun$ of molecular gas in the BCG in three large
filaments extending to $3-5\kpc$ radii.  The molecular cloud
velocities are remarkably low, within $\pm100\kmps$ of the BCG's
systemic velocity, and the velocity dispersion is significantly less
than the typical stellar velocity dispersion of such a massive BCG.
Apparently, the molecular gas has not settled in to the gravitational
potential well and the filament structure is expected to break up on
$<10^7\yr$ timescales unless it is supported, possibly by magnetic
fields.  The low velocities and shallow velocity gradients along the
filaments are inconsistent with free-fall or a merger.  The
molecular gas clouds must have originated from gas cooling less than a
few kpc from their current location or much higher velocities would be
observed.  However, although the X-ray cooling rate from the XMM-RGS
could supply $\sim5\times10^9\Msun$ of molecular gas on a $20\Myr$
timescale, such rapid cooling would dramatically deplete the hot gas
within the $\sim5\kpc$ radius of the filaments.


The N and SE filaments are projected beneath cavities in X-ray surface
brightness images corresponding to large radio bubbles inflated by the
central AGN that are now buoyantly rising through the cluster.  These
filaments contain $60-70\%$ of the total molecular gas mass, therefore
direct uplift by the radio bubbles appears to require an implausibly
high coupling efficiency.  In a cluster where AGN heating is locally
balanced by radiative cooling, lifting low entropy gas from its
equilibrium state at the cluster centre can promote cooling.  The cold
filaments are coincident with low temperature X-ray gas, bright
optical line emission and dust lanes suggesting that the molecular gas
could have formed by gas cooling from uplifted warmer gas.  The
survival of these extended cold gas filaments for at least a dynamical
timescale suggests magnetic fields are required for support and
insulation against the hot cluster atmosphere.  Supporting the weight
of the massive molecular filaments requires a magnetic pressure that
is an order of magnitude greater than the hot gas pressure.
\textit{Chandra} observations of PKS\,0745-191 are consistent with
departures from hydrostatic equilibrium in the cluster core and
non-thermal pressure contributions within a radius of $40\kpc$.


The N filament extends furthest and appears to be breaking into clumps
at large radius.  The filaments may eventually fragment with gas
clouds falling back onto the BCG centre in a circulation flow that
subsequently fuels the central AGN.  However, if the filaments are
formed from rapid cooling of the cluster atmosphere at
$\sim200\Msunpyr$, we would expect to detect molecular gas structures
associated with previous cooling episodes.  The prevalence of short
central radiative cooling times in cluster atmospheres and strong
correlations with detections of cold gas and star formation suggest
that cooling is long-lived and the AGN supplies a regular input of
energy.  Whilst a single episode of rapid cooling seems unlikely, it
also appears implausible that such a large fraction of the cold gas in
previous inflows could have been rapidly accreted whilst forming few
young stars.  We note that an $X_{\mathrm{CO}}$ factor for BCGs that
falls significantly below the Galactic value could alleviate several
problems, including the demands on magnetic pressure and depletion of
the hot gas.  However, the low gas velocities in PKS\,0745-191 and narrow
velocity dispersions observed for molecular clouds in BCGs are
inconsistent with the kinematic disturbances in starburst galaxies
that require significantly lower $X_{\mathrm{CO}}$ factors.  The low
metal abundance suggests that the molecular gas mass could instead
have been underestimated.



HST FUV observations of PKS\,0745-191 show a bright spur of emission from
young, massive star formation at $10-20\Msunpyr$ coincident with the
SW molecular gas filament.  The SW filament has a comparable velocity
structure and mass to the other two filaments and such a clumpy,
highly asymmetric distribution of gas about the nucleus does not
appear consistent with a gas disk.  The small-scale radio structure detected in
$14.9\GHz$ VLA radio observations is also spatially coincident with
the SW filament.  A collision between an expanding radio structure and the molecular gas may
have triggered collapse of the gas clouds and star formation in
the SW filament.  Disruption of the radio structure on small scales is also
consistent with the amorphous radio morphology on larger scales in low
frequency observations.  However, no increase in the velocity dispersion is observed either
along the SW filament or by comparison with the other filaments.

\section*{Acknowledgements}

HRR and ACF acknowledge support from ERC Advanced Grant Feedback 340442.  BRM
acknowledges support from the Natural Sciences and Engineering Council
of Canada and the Canadian Space Agency Space Science Enhancement
Program.  PEJN was partly supported by NASA contract NAS8-03060.  ACE
acknowledges support from STFC grant ST/L00075X/1.  GRT acknowledges support from Einstein Postdoctoral Fellowship Award Number PF-150128, issued by the Chandra X-ray Observatory Center, which is operated by the Smithsonian Astrophysical Observatory for and on behalf of NASA under contract NAS8-03060.  We thank the
reviewer for their thorough reading of the paper and encouraging
comments.  This paper makes use of the following ALMA data:
ADS/JAO.ALMA 2012.1.00837.S. ALMA is a partnership of ESO
(representing its member states), NSF (USA) and NINS (Japan), together
with NRC (Canada), NSC and ASIAA (Taiwan), and KASI (Republic of
Korea), in cooperation with the Republic of Chile. The Joint ALMA
Observatory is operated by ESO, AUI/NRAO and NAOJ.  The scientific
results reported in this article are based on data obtained from the
Chandra Data Archive.  This publication makes use of data products
from the Two Micron All Sky Survey, which is a joint project of the
University of Massachusetts and the Infrared Processing and Analysis
Center/California Institute of Technology, funded by the National
Aeronautics and Space Administration and the National Science
Foundation.

\bibliographystyle{mn2e} 
\bibliography{refs.bib}

\clearpage

\end{document}